\documentclass[11pt,tightenlines,eqsecnum,floats,aps,amsmath,amssymb,nofootinbib,prd,shownopacs,floatfix]{revtex4}
\usepackage{graphicx, wrapfig}
\usepackage{amssymb}
\usepackage{color}
\usepackage{mathrsfs}
\usepackage{subfigure}
\def\f{\frac}

\def\dd{\textrm{d}}

\def\t{\tilde}
\def\b{\bar}

\def\lp{l_{\rm Pl}}
\def\spl{s_{\rm Pl}}

\def\rhopl{\rho_{\rm Pl}}

\def\dphi{\partial_\phi}

\def\Q{{\mathcal Q}}
\def\t{\tilde}
\def\h{\hat}
\def\u{\mathfrak{A}}

\def\Nbstar{N_{{\rm B}\,\star}}

\def\phib{\phi_{\rm B}}
\def\rhob{\rho_{\rm B}}

\def\mpc{\rm Mpc^{-1}}

\def\Hp{\mathcal{H}_{\rm phy}}

\usepackage{enumerate}

\newcommand{\ig}{\includegraphics}
\newcommand{\be}{\nopagebreak[3]\begin{equation}}
\newcommand{\ee}{\end{equation}}
\newcommand{\bfig}{\nopagebreak[3]\begin{figure}}
\newcommand{\efig}{\end{figure}}
\newcommand{\ba}{\nopagebreak[3]\begin{eqnarray}}
\newcommand{\ea}{\end{eqnarray}}

\newcommand{\bmult}{\nopagebreak[3]\begin{multline}}
\newcommand{\emult}{\end{multline}}
\newcommand{\fref}[1]{Fig.\,\ref{#1}}

\def\aE{\bar{a}_{\rm E}}
\def\aGE{\bar{a}_{\rm GE}}
\def\at{\tilde{a}}
\def\etat{\tilde\eta}
\def\V{V_{\rm phy}}
\def\rhosup{\rho_{\rm sup}}
\def\gE{{\bar{g}^{\rm (E)}}}
\def\gGE{{\bar{g}^{\rm (GE)}}}

\def\rb{\t{r}_{\rm B}}
\def\rbL{\t{r}_{\rm B}^{(L)}}
\def\rbS{\t{r}_{\rm B}^{(S)}}
\def\lambdap{\lambda_{\rm phys}}

\begin{document}
\title{Phenomenology with fluctuating quantum geometries\\ in loop quantum 
cosmology\\
}
\author{Ivan Agullo$^1$}
\email{agullo@lsu.edu}
\author{Abhay Ashtekar$^2$}
\email{ashtekar@gravity.psu.edu}
\author{Brajesh Gupt$^2$}
\email{bgupt@gravity.psu.edu}

\affiliation{
$^1$
Department of Physics and Astronomy, Louisiana State University, Baton Rouge, LA 70803, U.S.A.
}
\affiliation{
$^2$ 
Institute for Gravitation and the Cosmos \& Physics Department, The Pennsylvania State University, University Park, PA 16802 U.S.A.
}

\begin{abstract}

The goal of this paper is to probe phenomenological implications  of large fluctuations of quantum geometry in the Planck era, using cosmology of the early universe. For the background (Friedmann, Lema\^{i}tre, Robertson, Walker)  \emph{quantum} geometry, we allow `widely spread' states in which the \emph{relative} dispersions are as large as $168\%$ in the Planck regime. By introducing suitable methods to overcome the ensuing conceptual and computational issues, we calculate the power spectrum $P_{\mathcal{R}}(k)$ and the spectral index $n_s(k)$ of primordial curvature perturbations. These results generalize the previous work in loop quantum cosmology which focused on those states which were known to remain sharply peaked throughout the Planck regime. Surprisingly, even though the fluctuations we now consider are large, their presence does not add new features to the final $P_{\mathcal{R}}(k)$ and $n_s(k)$: Within observational error bars, their effect is degenerate with a different freedom in the theory, namely the number of \emph{pre-inflationary} e-folds $\Nbstar$ between the bounce and the onset of inflation. Therefore, with regard to observational consequences, one can simulate the freedom in the choice of states with large fluctuations in the Planck era using the simpler, sharply peaked states, simply by allowing for different values of $\Nbstar$.

\end{abstract}

\maketitle

\section{Introduction}
\label{s1}

Over the past decade, loop quantum cosmology (LQC) has emerged as a promising candidate to describe the very early universe in a coherent and self-consistent manner (see, e.g., reviews \cite{as1,agullocorichi,abrev,reviewmena,grainrev}, and references therein). In this framework, quantum gravity effects are completely negligible away from the Planck regime, making general relativity an excellent approximation. However, once energy density and curvature approach the Planck scale, quantum geometry underlying LQC dominates. The big bang singularity is naturally resolved and replaced by a quantum bounce in all commonly used cosmological models. In the Planck epoch, cosmological perturbations now propagate on the Friedmann, Lema\^itre, Robertson, Walker (FLRW) \emph{quantum} geometry. Detailed analysis of this dynamics has provided viable extensions of the standard inflationary scenario over 11-12 orders of magnitude in curvature that bridge the onset of inflation with the Planck regime. Finally,  possible observational implications of this pre-inflationary dynamics have been studied in detail more recently \cite{aan1,aan2,aan3,am,agulloassym,ag3,mdbo2016}.

Most of this analysis was carried out in the context of spatially flat FLRW geometries. A quantum FLRW geometry is described by a wave function  $\Psi_{o}(v,\phi)$, where volume $v$ (related to the scale factor by $v \sim a^{3}$) represents the geometrical degree of freedom and $\phi$, the inflaton field.%
\footnote{The subscript in $\Psi_{o}$ is a reminder that it is the wave function of the \emph{background} geometry. If the spatial topology is $\mathbb{T}^{3}$, then $v$ is the physical volume of the universe. If it is $\mathbb{R}^{3}$, one introduces an infrared cut-off using finite box and $v$ is then the volume of the box. To extract physical results, one removes the cut-off by letting the box fill the whole universe at the end of calculations. In the classical theory, this procedure is not needed to write the field equations. However, the procedure is essential in order to cast the theory in a Hamiltonian or Lagrangian form which serve as the point of departure for the quantum theory. For details, see e.g.\cite{as1}.} 
Since quantum geometry effects become negligible quickly once the energy density and curvature fall well below the Planck scale, it turns out that one only needs the full LQC equations in the relatively short epoch around the bounce, which we will refer to as the \emph{Planck regime}. During this epoch, the field $\phi$ can be taken to be a relational time variable and the Hamiltonian constraint --the LQC analog of the Wheeler DeWitt equation-- on $\Psi_{o}(v,\phi)$ can be interpreted as the evolution equation in this internal time. 

Up to now, most of the phenomenological work in LQC makes a crucial use of the following fact. Consider the post-bounce, expanding branch of the universe. In this branch, one can assume that the quantum state $\Psi_{o}(v,\phi)$ is sharply peaked in $v$ \emph{in the low curvature regime} because quantum fluctuations in geometry are negligible during this epoch. It turns out that, under the LQC evolution \emph{backward in time}, such states remain sharply peaked all the way to the bounce (see, e.g., \cite{aps3}). Therefore phenomenological analyses often start with states $\Psi_{o}$ of the background FLRW geometry that are sharply peaked at the bounce and investigates the dynamics of quantum fields representing scalar and tensor modes on these quantum geometries. This  procedure is well-motivated. Nonetheless, since the LQC physical Hilbert space $\Hp$ does allow widely spread states, one can ask what would happen if one allows states $\Psi_{o}$ that have large relative dispersions in geometry in the Planck regime. Will such background quantum geometries introduce a variety of ambiguities in predictions of power spectrum and spectral index for scalar and tensor modes? Or, is there some underlying mechanism that prevents a proliferation of such ambiguities? How robust are the LQC predictions against this freedom? These questions allow us to probe LQC from a new angle and the ensuing answers should lead to a deeper insight into the structure of the theory.  

States with large dispersions in the Planck regime are not excluded a priori in the mainstream paradigm of the early universe since space-time geometry is assumed to be classical only when one is well away from the Planck epoch. In the standard inflationary scenario, for example, the curvature scale is $10^{-11} $-$10^{-12}$ times the Planck scale. However, findings in LQC discussed above do tell us that states with large relative dispersions will not display the desired classical behavior away from the Planck regime if their dynamics is dictated entirely by the Hamiltonian constraint. One would need a new mechanism to reduce quantum fluctuations and make the state sharply peaked on a classical trajectory once it enters the low curvature regime.%
\footnote{Contrary to common intuition, our results will show that it is possible that the \emph{physical} FLRW geometry could have large fluctuations even during inflation. The transition to classical geometry could then take place as late as the end of inflation. We will return to this point in section
\ref{s5}.}
We are not aware of a detailed mechanism that would be adequate. But a priori one cannot rule out the possibility that some novel ideas will emerge, inducing the desired classical behavior at late times even though the state is widely spread in the Planck regime. (An example in this direction is provided by Ref. \cite{stamp}, which proposes a generalization of quantum mechanics.) Therefore, in order to analyze the structure of LQC from a more demanding criterion, in this paper we will address the phenomenological questions raised above, assuming that the desired classical behavior does emerge in the appropriate regime via some such mechanism.

The phenomenological analysis using widely spread states is not straightforward because one faces difficult issues on both conceptual and computational fronts. On the conceptual side, one faces a fundamental question: How do you describe dynamics of test quantum fields representing cosmological perturbations on a background quantum geometry $\Psi_{o}$ with large fluctuations? For sharply peaked states one could imagine using the standard effective equations of LQC \cite{vt,aps3,as1} whose solutions trace the peak of the wave function $\Psi_{o}$. Although the smooth space-time metric provided by these solutions shows major deviations from Einstein dynamics, it is nonetheless a metric with FLRW symmetries and one knows how to analyze quantum fields propagating on general FLRW backgrounds. If the states have wide relative dispersions, the derivation \cite{vt} of these effective equations breaks down, and effective equations fail to capture important qualitative features of LQC dynamics in the Planck regime \cite{ag1}. On the computational side, new challenges emerge already at the level of the background dynamics! Methods that were used to analyze the sharply peaked states \cite{aps3} are no longer adequate and significantly more sophisticated techniques are now needed \cite{dgs1,dgs2,dgms}. Furthermore, to evolve cosmological perturbations, one needs to calculate expectation values of certain operators in the background quantum geometry $\Psi_{o}$ and this task is also computationally non-trivial \cite{ag1}.

The paper is organized as follows. In order to have some confidence in the robustness of the final picture, in the detailed analysis we consider two classes of widely spread states --Gaussian and non-Gaussian with multiple peaks-- and two distinct inflation potentials --the quadratic and the Starobinsky. In section \ref{s2}, we recall the relevant results on dynamics of the background  FLRW quantum geometry and of cosmological perturbations propagating on this background, emphasizing the new features that arise when sharply peaked states are replaced by those with large relative dispersions in geometry. It turns out that the conceptual problem mentioned above can be overcome using the `dressed effective metric' strategy introduced in \cite{akl} so long as the back reaction of perturbations on the background quantum FLRW geometry can be ignored. In section \ref{s3} we undertake the task of computing the quantum corrected, dressed effective metric which governs the dynamics of perturbations, under the assumption that the potential energy at the bounce is negligible compared to the kinetic energy.
This task is numerically challenging and requires us to restrict the class of widely spread states $\Psi_{o}$. We do allow relative dispersions up to $168\%$ in the Planck regime but, even with improved schemes introduced in \cite{dgs1}, computational limitations do not allow us to go further. In sections \ref{s4} we investigate phenomenological implications of states with large fluctuations in geometry. The surprising finding is that in spite of relative dispersions of $168\%$ in the Planck regime ,  there are \emph{no new features in the scalar power spectrum}. Each state $\Psi_{o}$ --sharply peaked or widely spread-- leads to pre-inflationary dynamics with a certain number $\Nbstar$ of pre-inflationary e-folds and, within observational error bars, \emph{the scalar power spectrum in the quantum geometry defined by a given widely spread state is identical to that defined by a sharply peaked state with a slightly greater value of $\Nbstar$.} Therefore, as far as power spectrum is concerned, one can restrict oneself to sharply peaked states corresponding to various values of $\Nbstar$. In section \ref{s5} we collect the assumptions that underlie this analysis, summarize the main results, and present arguments that provide a conceptual understanding of the main results.

Note that, throughout this paper, terms `fluctuations' and `dispersions' refer to the quantum state of the \emph{homogeneous} background geometries; \emph{not} to the cosmological perturbations which are inhomogeneous.

\section{Quantum FLRW Geometries}
\label{s2}

To make the paper self-contained, in this section  we recall the relevant properties of the FLRW quantum geometries $\Psi_{o}(v,\phi)$, spell out the sense in which effective descriptions can be used to approximate their quantum dynamics, and discuss the propagation of cosmological perturbations on these  geometries. This summary will be brief. For details and subtleties, see for examples the papers \cite{aps1,aps2,aps3,acs,akl,aan2,aan3} where there results first appeared, or review articles, e.g., \cite{as1,agullocorichi,abrev}. We will also specify the type of widely spread states that will be used in our numerical simulations in the rest of the paper.

\subsection{Physical states $\Psi_{o}(v,\phi)$}
\label{s2.1}

Consider the mini-superspace consisting of spatially flat FLRW geometries with $\mathbb{R}^{3}$ spatial topologies, where the only matter is a scalar field $\phi$. In LQC, one typically uses the Hamiltonian framework to pass to the quantum theory. As explained in footnote 1, this requires one to introduce a box, say of comoving volume $V_{o}$, as an infrared regulator, carry out the analysis and remove the regulator by taking the limit $V_{o} \to \infty$ at the end. The gravitational phase space is then spanned by the canonically conjugate pair $(v,h)$. The geometrical meaning of these variables is the following. The \emph{physical} volume $\V$ of the box is given by 
\be \V \equiv a^{3}V_{o} = 2\pi\, G v\, . \ee
Thus, $a = (V_o^{-1} 2\pi G v)^{1/3}$ is the scale factor, and on dynamical trajectories representing solutions to Einstein's equations $h$ turns out to be the Hubble parameter. (For further details, see \cite{ag1}. We will use the notation introduced there.) The matter part of the phase space is spanned by the canonically conjugate pair $(\phi,p_{\phi})$. The kinematical Hilbert space --the quantum analog of the full phase space-- consists of states $\Psi_{o}(v,\phi)$. As usual, dynamics is governed by the Hamiltonian constraint. In LQC, this constraint is given by \cite{ach2}
\footnote{The variable $\nu$ used in \cite{ach2} is related to our $v$ via $v= \gamma \hbar \nu$ where $\gamma \equiv \Delta_{o}/4\sqrt{3}\pi$ is the Barbero-Immirzi parameter. As in \cite{ag1}, our viewpoint is that it is the area-gap that is the new \emph{physical} parameter of the theory and $\gamma$ is a mathematical parameter introduced in the passage from classical general relativity to LQG. Therefore, only $\Delta_{o}$ will feature in all our expressions. As with $\nu$, positive values of our $v$ refer to the post-bounce branch and negative values to the pre-bounce branch. In this paper we are interested only in the post-bounce phase. Finally, while in \cite{ach2} $\ell_{o}^{2}$ is set equal to the area gap, here $\ell_{o}^{2}= {\Delta^3_o}/{48\pi^2}\lp^2$.} 
\ba
  \hbar^{2}\dphi^2\Psi_o(v,~\phi) &=& \f{3\pi G}{4\ell_o^2} \Big[ \sqrt{v(v+4\hbar\ell_o)}(v+2\hbar\ell_o)\, \Psi_o(v+4\hbar\ell_o,\phi) - 2 v^2\, \Psi_o(v,\phi)  \nonumber\\ 
                     &&  + \sqrt{v(v-4\hbar\ell_o)}(v-2\hbar\ell_o)\, \Psi_o(v-4\hbar\ell_o)\Big]\, +\, \Big[8\pi^{2}G^{2} v^{2} V(\phi)\, \Psi_{o}(v,\phi)\Big] \nonumber \\
       &=:& - \big(\Theta_{0} + \Theta_{1}\big)\, \Psi_o(v,~\phi) \equiv - \Theta\, \Psi_o(v,~\phi).
\label{eq:lqchamilt}
\ea
Here $V(\phi)$ is the inflaton potential, $\lp$ the Planck length, and $\ell_o^2$ is related to the \emph{area gap} $\Delta_{o}$, the minimum non-zero eigenvalue of the area operator, via $\ell_{o}^{2}= ({\Delta^3_o}/{48\pi^2})\,\lp^2$. The inflaton potential enters only the operator $\Theta_{1} = 8\pi^{2}G^{2} v^{2} V(\phi)$ on the right side. The operator $\Theta_{0}$ is the geometric part of the total operator $\Theta$. It is a second order \emph{difference} operator with step size $4\hbar\ell_{o}$, determined by the area gap. If kinetic energy of the inflaton dominates at the bounce, then in the Planck regime $\Theta_{0}$ can be regarded as the main part of $\Theta$ and $\Theta_{1}$ can be regarded as a perturbation. A systematic calculation shows that, in the limit in which the area gap goes to zero and quantum geometry underlying Loop Quantum Gravity (LQG) is ignored, $\Theta_{0}$ is well-approximated by differential operator and (\ref{eq:lqchamilt}) reduces to the Wheeler-DeWitt equation. 

Suitably regular solutions $\Psi_{o}$ to (\ref{eq:lqchamilt}) represent physical states of the FLRW quantum geometry. More precisely, the physical Hilbert space $\Hp$ consists of solutions to (\ref{eq:lqchamilt}) that are normalizable with respect to an appropriate inner product. The so-called `group averaging procedure' provides a natural avenue to complete this task \cite{dm,abc}. This is a step by step, constructive procedure that is applicable for a large class of constrained systems. However, the resulting inner product can be non-local in the $v,\phi$ configuration space. In that case, it is difficult to construct interesting Dirac observables and extract the physical content of the theory explicitly. In some cases, e.g. when the potential $V(\phi)$ vanishes, or corresponds just to a cosmological constant, the procedure is known to lead to a manageable inner product that is simple enough to enable one to easily extract physics (both in LQC and the Wheeler-DeWitt theory). We will summarize the structure of the resulting theory for the case when $V(\phi)=0$, i.e., where $\Theta=\Theta_{0}$ in (\ref{eq:lqchamilt}). (For reasons explained in section \ref{s3}, this special case plays a key role in our analysis.)

When $\Theta=\Theta_{0}$, the total Hilbert space of all normalizable solutions has two `superselection' sectors, consisting of solutions to (\ref{eq:lqchamilt}) of the type
\be \label{evo} \mp i\hbar\dphi \Psi_{o}(v,\phi) = \sqrt{\Theta_{0}}\, \Psi_{o}(v,\phi) \equiv H_{o} \Psi_{o} (v,\phi)\, . \ee
These are the analogs of the positive and negative frequency sectors of the familiar example of Klein-Gordon equation in stationary space-times. It suffices to restrict oneself just to one of them, say the `positive frequency' sector with the upper sign. Then the Physical Hilbert space $\Hp$ consists of all solutions to the `Schr\"odinger like' equation $-i\hbar \dphi \Psi_{o}(v,\phi) = H_{o} \Psi_{o} (v,\phi)$ with finite norm:
\be \label{norm} ||\Psi_{o}||^{2} \, :=\, \sum_{v} |\Psi_{o}(v,\phi)|^{2} \, <\, \infty \, .\ee 
Note that: (i) in virtue of the right side of (\ref{eq:lqchamilt}), the sum is restricted to $v = 4n\hbar\ell_{o}$ where $n$ runs over positive integers; and (ii) the norm can be computed at any value of $\phi$; it is independent of the choice of $\phi$ because of the `Schr\"odinger equation' satisfied by $\Psi_{o}(v,\phi)$.

Each state $\Psi_{o}(v,\phi) \in \Hp$ is the analog of a dynamical trajectory in the classical phase space and thus represents a quantum FLRW geometry. In this description it is convenient --although not essential-- to think of $\phi$ as representing a relational time variable with respect to which the physical volume $v$ of the box --or, equivalently, the scale factor-- `evolves'. The volume of the box at any given value $\phi_{0}$ of the internal time is  a Dirac observable, represented by the operator $\h{V}\mid_{\phi_{0}}$ given by
%
%
\be  \h{V}\mid_{\phi_{0}}\, \Psi_{o}(v,\phi) = e^{\f{i}{\hbar} H_{o}(\phi-\phi_{0})}\, (2\pi G v)\, \Psi_{o} (v,\phi_{0}) \, . \ee
Thus, to operate by $\h{V}\mid_{\phi_{0}}$ on any physical state $\Psi_{o}(v,\phi) \in \Hp$, one freezes that state at $\phi=\phi_{0}$, acts on it by the kinematical volume operator $\h{V}$ and evolves the result using $-i\hbar \dphi \Psi_{o}(v,\phi) = H_{o} \Psi_{o} (v,\phi)$ to obtain a new physical state. The operators  $\h{V}\mid_{\phi_{0}}$ refer to the box used in the infrared regularization. However, the scale factor and matter density operators operators defined via
\be \label{ahat} \Big(\h{a}_{\mid_{\phi_{0}}}\Big)^{3} :=  V_{o}^{-1}\, \h{V}\mid_{\phi_{0}} 
\quad {\rm and} \quad   \h\rho_{\mid_{\phi_{0}}} :=\f{1}{2}\, \big(\h{V}\mid_{\phi=\phi_{0}}\big)^{-1}\,\, \h{p}_{\phi}^{2}\,\, \big(\h{V}\mid_{\phi=\phi_{0}}\big)^{-1}\, ,\ee 
do not depend on the size of the box. Therefore, in statements regarding their dynamics, one can trivially remove the infrared regulator.

Given any state $\Psi_{o}$ in (a dense subspace of) $\Hp$, one can show that during its `evolution', there is precisely one `instant' $\phib$ of the relational time $\phi$ at which the expectation value $\h{V}\mid_{\phi_{\rm B}}$ attains its minimum, $2\pi G v_{\rm min}$, and the expectation value of $\h\rho\mid_{\phi}$ attains its maximum $\rho_{\rm max}$. This is the `instant' at which the universe in that state bounces. One can show that the eigenvalues of operators $\h\rho\mid_{\phi_{0}}$ have an \emph{absolute supremum} on $\Hp$ with 
\be \rhosup = \f{18\pi}{G^2\hbar\Delta_o^3} \approx 0.4092~\rhopl\, ,\ee 
where we use the value $\Delta_o= 5.17$ from black hole entropy calculations given in \cite{meissner,abbdv}. 
If the state $\Psi_{o}$ is sharply peaked in volume at the bounce, the expectation value of $\h\rho\mid_{\phib}$ is very close to $\rhosup$; narrower the relative dispersion in volume at the bounce, closer is the expectation value to $\rhosup$. On the other hand, if the state has wide dispersion, $\rhob$ can be much lower than $\rhosup$. The precise relation between the relative dispersion in the volume in a state $\Psi_{o}$ and the density $\rhob$ at the bounce is discussed in Ref. \cite{ag1}.

\bfig
 \ig[width=0.45\textwidth]{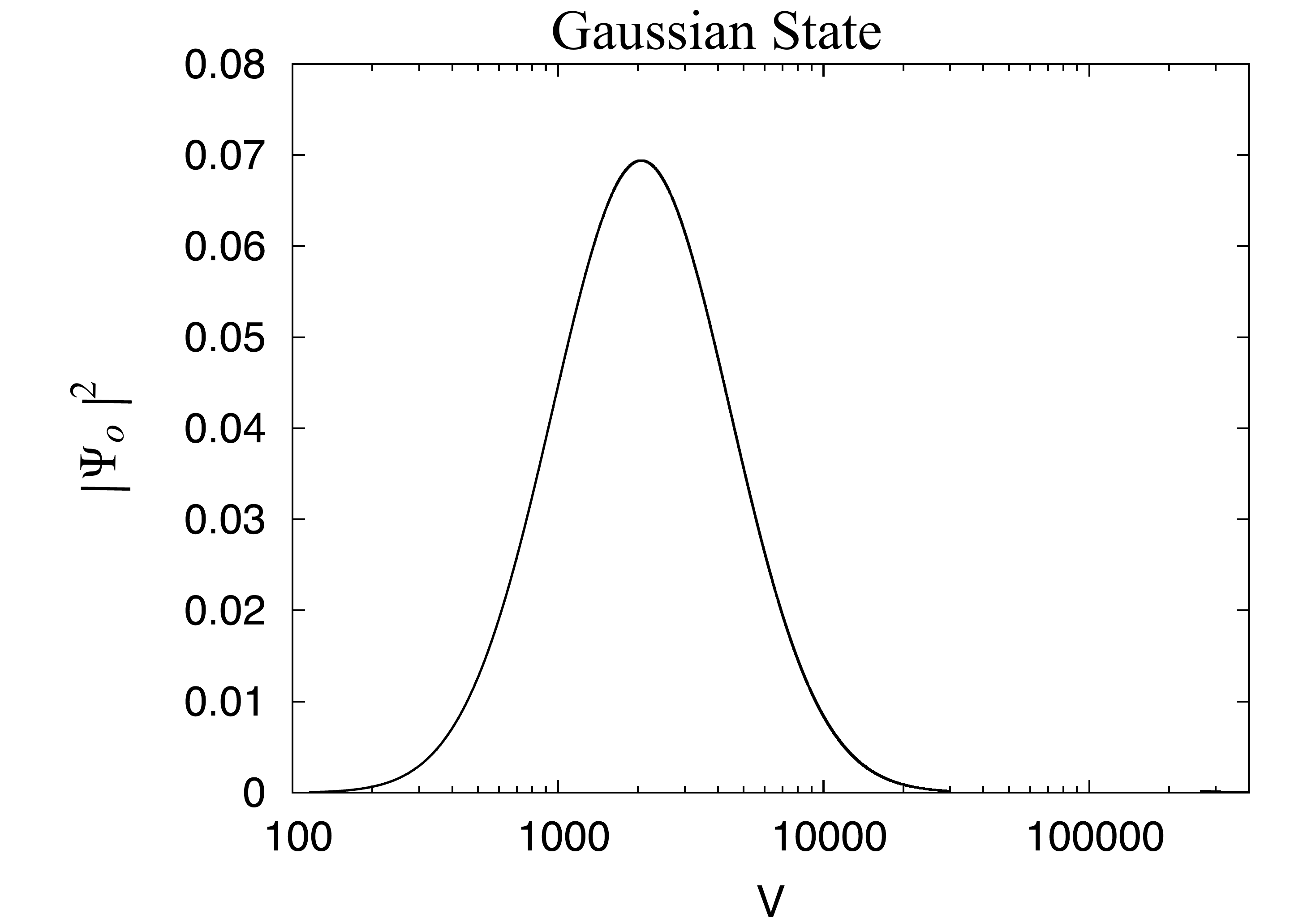}
\hskip0.2cm
 \ig[width=0.45\textwidth]{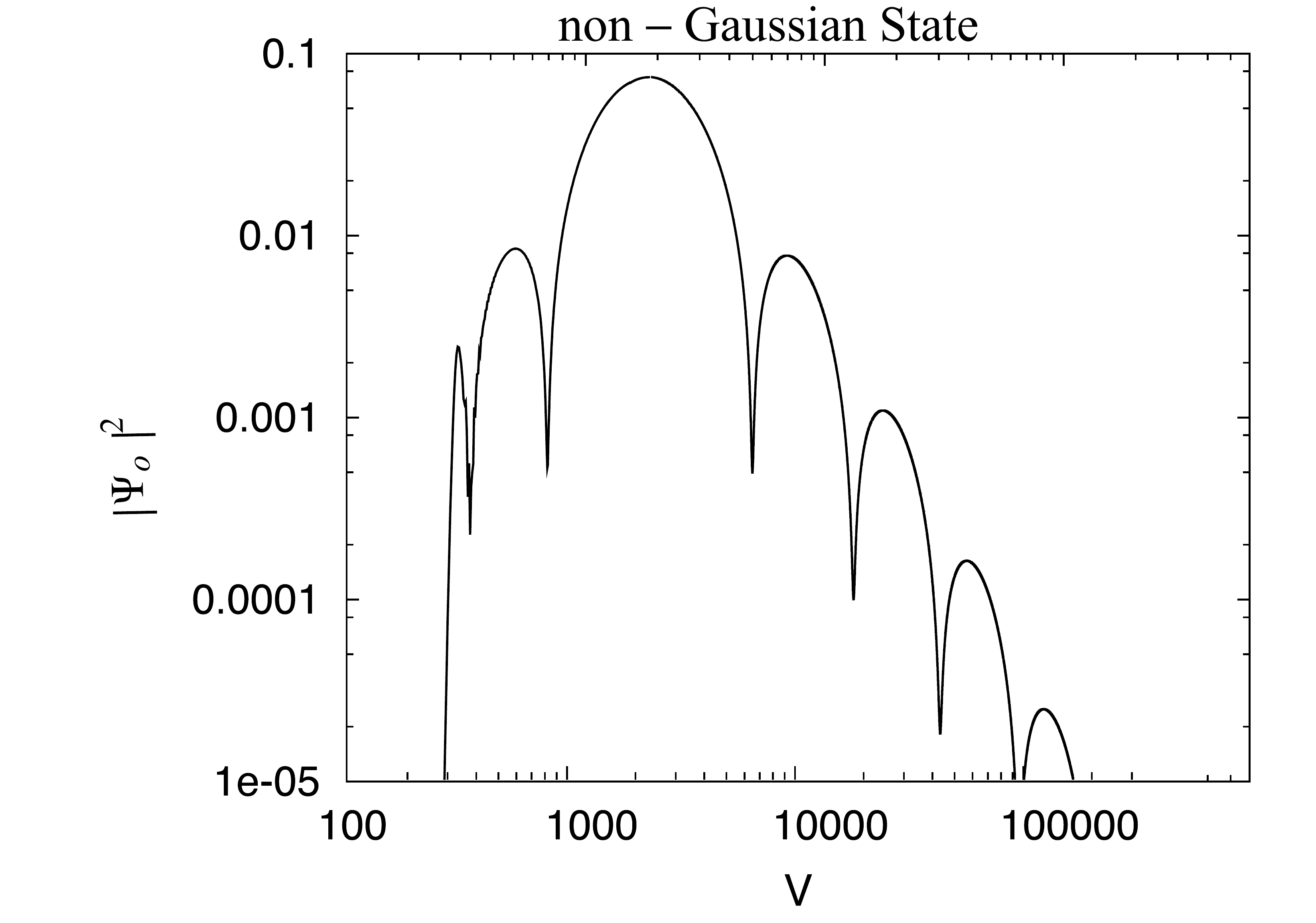}
\caption{Examples of widely spread states used in simulations. 
\emph{Left Panel:} A Gaussian wave function; and, \emph{Right Panel:} A multi-peaked, non-Gaussian wave function. In simulations both types of states will have $100\%$ relative dispersion in volume at the bounce. The dispersion increases away from the bounce and saturates at approximately $168\%$. We will compare their dynamics with that of sharply peaked Gaussians which have a relative dispersion of $1\%$ at the bounce.}
\label{fig:wavefun}
\efig

In this paper we will consider two families of states, whose form at the bounce is shown in Fig. \ref{fig:wavefun}: Gaussian (left panel) and non-Gaussian states with multiple peaks (right panel). To explicitly construct them, we first find eigenfunctions of the difference operator $\Theta_{0}$ that features in (\ref{eq:lqchamilt}): 
\be
 \Theta_{0}\, e_{\underbar k}(v) = \hbar^2\omega({\underbar k})^2~e_{\underbar k}(v) \qquad {\rm with} \qquad
\omega({\underbar k}) = \sqrt{12\pi G}~|{\underbar k}|; \quad -\infty<{\underbar k}<\infty\, .
\ee 
As in previous works \cite{aps2,aps3,dgs2,dgms} these calculations will be performed numerically. (Although an analytical expression of the eigenfunctions $e_{\underbar k}(v)$ is available \cite{ach2}, since it involves an integral over $h$, the momentum conjugate to $v$, numerically it is easier to solve for eigenfunctions and eigenvalues directly.) Then, the wave function $\Psi_{o}(v, \phib)$ can be expanded out in terms of these eigenfunctions;
\be \label{state} \Psi_{o}(v,\phib) = \int_{-\infty}^{\infty}\!\! \dd \underbar k \,\t{\Psi}(\underbar k)\, e_{\underbar k}(v)\, e^{i\omega\phib} . \ee
Therefore, to specify $\Psi_{o}(v,\phib)$ it suffices to choose a profile $\t\Psi(\underbar k)$. By `Gaussian state' we mean a quantum geometry state $\Psi_{o}(v,\phib)$ at the bounce whose profile has the form 
\be
\tilde\Psi(\underbar k) = \exp{[-\f{(\underbar k-\underbar k_o)^2}{2\sigma^2}]},
\label{eq:gauss}
\ee
that is determined by the choice of two parameters, $\underbar k_{o}$ and $\sigma$. In the early papers on LQC, the parameters were so chosen that $\Psi_{o}(v,\phib)$ is sharply peaked in $v$, in the sense that the relative dispersions in $\h{V}\mid_{\phib}$ are very small. In this paper, following \cite{dgs2,dgms} we will consider widely spread states, now allowing relative dispersions of $100\%$. The non-Gaussian states we consider have the profile function $\t \Psi(\underbar k)$ of the form \cite{dgms}
\be
\t\Psi(\underbar k) = \exp{\left[-\f{\left(\underbar k-\left(\underbar k_o + \delta \underbar k \right)\right)^2 
                     \left(\underbar k-\left(\underbar k_o - \delta \underbar k \right)\right)^2}{\sigma^4}\right]},
\label{eq:nongauss}
\ee
parametrized by $\underbar {k}_{o},\,\sigma$ and $\delta \underbar k$. As the right panel of Fig. \ref{fig:wavefun} shows, these states have multi-peaks. Again the parameters will be chosen to allow for relative dispersions in $v$ that are as large as $100\%$.
 
\subsection{Effective equations}
 \label{s2.2}

Let us begin with sharply peaked states $\Psi_{o}(v,\phi)$. In this case, the framework of geometrical quantum mechanics \cite{aats} allows one to determine the `trajectory' followed by the peak of the wave function \cite{vt,as1}. Furthermore, from this trajectory one can construct an effective space-time metric $\gE_{ab}$ around which quantum geometry is sharply peaked:
\be \gE_{ab} \dd x^{a} \dd x^{b} = - \dd t^{2}\, +\, \aE^{2}\,\, \dd \vec{x}^{2}\, . \ee
Away from the Planck regime, the effective scale factor $\aE$ satisfies Einstein's equations. However, in the Planck regime, there are large deviations from general relativity and the scale factor exhibits the bounce. More precisely, the Friedmann and Raychaudhuri equations are modified 
\ba
\label{EE}   H^2 &:=& \Big(\,\f{{\dot{\bar{a}}_{\rm E}}}{\aE}\,\Big)^{2}\,=\, \f{8\pi G}{3}\rho\, \Big(1-\f{\rho}{\rhosup}\Big) \nonumber\\
\dot H &=& -4\pi G \Big(\rho + P\Big)\,  \Big(1-2 \f{\rho}{\rhosup}\Big),
\ea
Here  $\rho(t)$ and $P(t)$ are the energy density and pressure of the scalar field and the dot indicates derivative with respect to cosmic time. Although these \emph{effective equations} (EE) capture only the leading order quantum corrections, they accurately track the evolution of the sharply peaked wave function $\Psi_{o}(v,\phi)$ also in the Planck regime, showing that the bounce occurs when the matter density $\rho$ achieves its maximum possible value $\rhosup$.

Let us now turn to states with large relative dispersions in the volume at the bounce. In this case, there is no obvious peak in $\Psi_{o} (v,\phi)$ whose evolution would give a quantum corrected trajectory. Furthermore the assumptions that were made in deriving effective equations (\ref{EE}) no longer hold and numerical simulations show that dynamics of the quantum state is \emph{no longer well-approximated by} (\ref{EE}) \cite{dgms}. Therefore, at first it would seem that there is no useful trajectory in the phase space that can be extracted from dynamics of these wave functions. Recall however that a \emph{general} state $\Psi_{o} \in \Hp$ does bounce in the sense that the \emph{expectation value} of the volume operators $\h{V}\mid_{\phi}$ in this state achieves the minimum and that of the density operators $\h\rho\mid_{\phi}$ attain the maximum at some value of $\phi$ which we denote by $\phib$ and call `the bounce time'. This suggests that a simple `evolution equation' could be extracted from the  \emph{expectation values} of various operators. While one would anticipate that some result along these lines should hold because of Ehrenfest's theorem in quantum mechanics several subtleties arise, first because definitions of quantities such as the matter density involve products of non-commuting operators, and second because one has to convert relational time to an appropriate proper time. As explained in section \ref{s3}, in this paper we will restrict ourselves to the case in which the potential $V(\phi)$ is subdominant at the bounce, and we will need the full LQC evolution only in the Planck regime, which lasts only $\sim 12$ Planck seconds. As we will see, for the phenomenological considerations on which the paper focuses, during this short epoch one can ignore the presence of the potential to an excellent degree of approximation. In this case, one can address the subtleties mentioned above and obtain equations governing dynamics of the mean scale factor. Surprisingly, these \emph{generalized effective equations} (GEEs) turn out to be remarkably simple extensions of (\ref{EE})\, \cite{ag1}:
\ba
\label{GEE}   {\bar{H}}^2 &:=& \Big(\,\f{{\dot{\bar{a}}_{\rm GE}}}{\aGE}\,\Big)^{2}\,=\, \f{8\pi G}{3}\bar{\rho} \Big(1-\f{\bar{\rho}}{\rhob}\Big) \nonumber\\
\dot{\bar{H}} &=& -4\pi G \Big(\bar{\rho} + \bar{P}\Big)\,  \Big(1-2 \f{\bar\rho}{\rhob}\Big)\, ,
\ea
where we have put a bar over $H, \rho$ and $p$ to emphasize that the equations only feature \emph{mean values} of appropriate operators in spite of the fact that $\Psi_{o}$ allows large fluctuations. Thus, rather surprisingly the generalization from sharply peaked states to the ones with wide dispersions consists only of replacing $\rhosup$, by $\rhob$. As remarked in section \ref{s2.1}, larger the relative dispersion in volume at the bounce, smaller is the density $\rhob$ at the bounce. But for sharply peaked states, $\rhob$ is very close to $\rhosup$ and Eqs. (\ref{GEE}) that hold for general states $\Psi_{o}$ reduce to Eqs. (\ref{EE}). 

To summarize, dynamics of the mean value of the scale factor is governed by rather simple GEEs under assumptions we make in this paper. Of course, when $\Psi_{o}$ is widely spread, the mean values capture a \emph{very} small part of the rich information contained in the dynamics of the full wave function $\Psi_{o}$. Nonetheless, as we will see in section \ref{s3}, even when the relative dispersions in volume are as large as $100\%$ at the bounce (and they \emph{grow} and plateau around $168\%$ as we evolve away from the bounce \cite{ag1}), the generalized effective trajectories satisfy Einstein's equations to an excellent degree of approximation just $\lesssim 5$ Planck seconds after the bounce! However quantum geometry effects during this short epoch are important as they \emph{can} leave observables imprints \cite{aan1,aan2,am,agulloassym}. The issue is whether these imprints can distinguish between states of quantum geometry that are sharply peaked near the bounce from those which have large fluctuations. To answer this question, in section \ref{s3} we will go beyond GEEs and use the quantum Hamiltonian constraint (\ref{eq:lqchamilt}).

\subsection{Quantum fields on the quantum geometry $\Psi_{o}(v,\phi)$}
\label{s2.3}

As in standard cosmology, in LQC cosmological inhomogeneous perturbations are described by 3 gauge invariant quantum fields $\h{\Q}(x)$, $\h{\mathcal{T}}^{(+)}(x)$, $\h{\mathcal{T}}^{(\times)}(x)$. They describe scalar curvature perturbations%
\footnote{In general relativity, the gauge invariant field $\h{\Q}(x)$ relates to the comoving curvature perturbations  $\h{\mathcal{R}}(x)$ commonly used in inflation by $\h{\mathcal{R}}= ({H}/{\dot \phi})\,\h{\Q}$. In the pre-inflationary evolution it is more convenient to work with $\h{Q}$, since $\h{\mathcal{R}}$ is not well-defined whenever $\dot \phi$ vanishes. The strategy is therefore to work with $\h{\Q}$ and translate the results to $\h{\mathcal{R}}$ at the end of inflation.} 
and the two polarization of gravitational waves, respectively. In this paper we focus only on the scalar modes. The main difference with standard cosmology is that in LQC these fields propagate in a \emph{quantum} FLRW spacetime described by $\Psi_o(v,\phi)$, rather than in a classical solution to Einstein's equations. The problem of propagating quantum fields in a quantum spacetime seems intractable at
first sight. However, the necessary conceptual framework was introduced in \cite{akl} and further developed for cosmological perturbations in \cite{aan2}. A key simplification occurs in cases when the back-reaction of the cosmological perturbations $\h{\Q}$ on the homogeneous background quantum geometry $\Psi_o(v,\phi)$ can be neglected. This is in fact an underlying assumption in most exploration of the early universe, such as inflation. Under these circumstances, the evolution of the background $\Psi_o(v,\phi)$ is unaltered by the presence of perturbations, and one can therefore solve for it before perturbations are introduced. Furthermore, there are unforeseen simplifications in the evolution of perturbations. First, explicit computations  \cite{akl,aan1,aan2,aan3} show that under these circumstances the evolution of scalar (and tensor) perturbations on the quantum geometry $\Psi_o(v,\phi)$ can be described by partial differential equations --rather than difference equations--  with time dependent coefficients. Second, these coefficients are obtained as expectation values of operators associated to the background geometry (certain combinations of $\hat a$ and $\hat H_0:=\sqrt{ \Theta_0}$). Third, the evolution equations have the \emph{same form} as in standard cosmology,
\be \label{eqnspert} (\t\Box - { \t{\u}})\,{\h{\mathcal{Q}}}(\vec{x},\t\eta) =0  \, \, , \hspace{1cm} 
\ee
where, however, the operator  $\t\Box:=\tilde{g}^{ab}\t\nabla_a \t\nabla_b$  and the potential $\t\u$ are now constructed using the state $\Psi_{0}(v,\phi)$ of the quantum FLRW geometry. More precisely, $\t\Box$
is the D'Alembertian associated to a smooth metric tensor with FLRW symmetries:
\be \label{dressedg}\tilde g_{ab}\dd x^a \dd x^b = \tilde{a}^2(\tilde{\eta})\,(-\dd\etat^2+\dd\vec{x}^2)\, , \ee
where the  scale factor $\t{a}$ is given by 
\be
\label{dresseda}  \tilde a^4 = \f{\langle \hat H_0^{-1/2}\hat{a}^4 \hat H_0^{-1/2}
\rangle} {\langle \hat  H_0^{-1} \rangle} \, , \ee
and the  conformal time $\tilde \eta$ is related to the internal time $\phi$ of LQC via
\be \label{dressedeta} \dd\tilde\eta = V_o \, (\langle H_0^{-1} \rangle)^{1/2} \,\, ( \langle H_0^{-1/2}
\hat{a}^4 H_0^{-1/2}\rangle)^{1/2} \, \dd\phi \, .
\ee
All expectations values are evaluated in the state $\Psi_{o}(v,\phi)$
chosen to describe the background quantum geometry. It is the only `free' Hamiltonian $H_{0}$ that features in these expressions because these calculations are carried out in the interaction picture. Finally, the potential  $\t{\u}(\tilde \eta )$ is given by 

\be \label{qpot} \t{\u} = \f{\langle \h{H}_o^{-\f{1}{2}}\,
\h{a}^2\, \h{\u}\,  \h{a}^2\, \h{H}_o^{-\f{1}{2}}
\rangle}{\langle \hat{H}_o^{-\f{1}{2}}\, \hat{a}^4\,
\hat{H}_o^{-\f{1}{2}}\rangle}\, . \ee
Here $\h{\u}$ is the operator corresponding to the classical potential ${\u}=a^2 [V(\phi)\,r - 2V_\phi(\phi)\sqrt{r} + V_{\phi\phi}(\phi)]$, where  $r=3 \dot \phi^2\,\frac{8\pi G}{\rho}$ is the fraction of the total energy density of the scalar field that is in the kinetic part;
$V(\phi)$ is the inflaton potential; $V_{\phi}(\phi)\equiv \dd V(\phi)/\dd\phi$; and $V_{\phi\phi}(\phi)\equiv \dd^2 V(\phi)/\dd\phi^2$. 

Note that \emph{if} the state were to be infinitely peaked at a classical FLRW solution $(g_{ab}, \phi)$ to Einstein-scalar field equation with \emph{zero} dispersions, then the complicated  expressions of $\t{a},\, \t\eta, \t\u$ would collapse to the scale factor, conformal time and the potential used in standard inflation in that solution. Of course the physical Hilbert space $\Hp$ does not admit such states that are infinitely peaked. Expressions (\ref{dresseda})-(\ref{qpot}) are complicated precisely because the state $\Psi_{o}(v)$ has fluctuations.
Thus, the scale factor $\t{a}$, the conformal time $\t\eta$ and the potential $\t\u$ know not only about the mean values of various quantities in the state $\Psi_{o}$ but also about a few specific fluctuations that are relevant to the right sides of Eqs. (\ref{dresseda})-(\ref{qpot}). Therefore, $\t{g}_{ab}$ is called the \emph{dressed effective metric} and $\t\u$ is called the \emph{dressed potential}. Both are smooth fields but depend on $\hbar$ in a subtle way, encoding certain aspects of the FLRW quantum geometry $\Psi_{o}(v,\phi)$. As is clear from explicit expressions of $\t{a},\t{\eta}$ and $\t\u$, the specific `dressing' that is needed to encode the effects of quantum geometry on dynamics of perturbations could not have been guessed a priori. It emerged from detailed calculations. 

Finally, note that these results do not require that the state $\Psi_{o}$ be sharply peaked. Thus, even when $\Psi_{o}(v,\phi)$ has large fluctuations with relative dispersions of $100\%$, cosmological perturbations $\h{\Q}(\vec{x},\t\eta)$, $\h{\mathcal{T}}^{(+)}(\vec{x},\t\eta)$, $\h{\mathcal{T}}^{(\times)}(\vec{x},\t\eta)$ are blind to most aspects of these fluctuations (so long as the back reaction of these perturbations on $\Psi_{o}$ is negligible). They are sensitive only to the three quantities $\t{a}, \t{\eta}$ and $\t{\u}$ extracted from $\Psi_{o}$. Put differently, if two states of quantum FLRW geometry with very different quantum fluctuations were to define the same $\t{a}, \t{\eta}$ and $\t{\u}$, we would not be able to distinguish between them using cosmological perturbations. However, the dressed effective metric $\t{g}_{ab}$ \emph{is} sensitive to certain fluctuations in $\Psi_{o}(v,\phi)$. Therefore,  the question is whether the $\t{g}_{ab}$ constructed from widely spread states are sufficiently different from those constructed from sharply peaked states to introduce new features in the dynamics of cosmological perturbations that would leave imprints on observational predictions.\\

\emph{Remark:} If the state $\Psi_{o}(v,\phi)$ is sharply peaked, the dressed effective metric $\t{g}_{ab}$ is well approximated by the effective metric $\b{g}_{ab}^{(E)}$ we discussed in section \ref{s2.2}. Therefore in the LQC literature often uses the effective metric while studying cosmological perturbations. We will see in section \ref{s3} that this  procedure is \emph{not} justified for states with large relative dispersions.

\section{The dressed metric $\t{g}_{ab}$}
\label{s3}

To calculate the power spectrum and the spectral index in the case when the state $\Psi_{o}$ has large fluctuations in geometry, we need to proceed in the following steps:\\
(i) Solve Eq. (\ref{eq:lqchamilt}) for  $\Psi_o(v,\phi)$ with appropriate initial data;\\
(ii) Compute the dressed effective scale factor $\t{a}$, conformal time $\t\eta$ and potential $\t\u$ using Eqs. (\ref{dresseda})-(\ref{qpot});\\
(iii) Solve  Eq. (\ref{eqnspert}) for the scalar mode $\h{Q}$ propagating on the dressed effective geometry $\t{g}_{ab}$; and,\\
(iv) Calculate the power spectrum $\mathcal{P}_{\mathcal{R}}(k)$ for curvature perturbations and the associated spectral index $n_{S}(k)$.\\
In this section we will carry out the first two steps and in section \ref{s4} the last two. 
In the first part of this section, we will introduce suitable approximations to solve (\ref{eq:lqchamilt}), and in the second we will calculate the dressed effective metrics $\t{g}_{ab}$ from the resulting wave functions $\Psi_{o}(v,\phi)$.

\bfig
 \ig[width=0.7\textwidth]{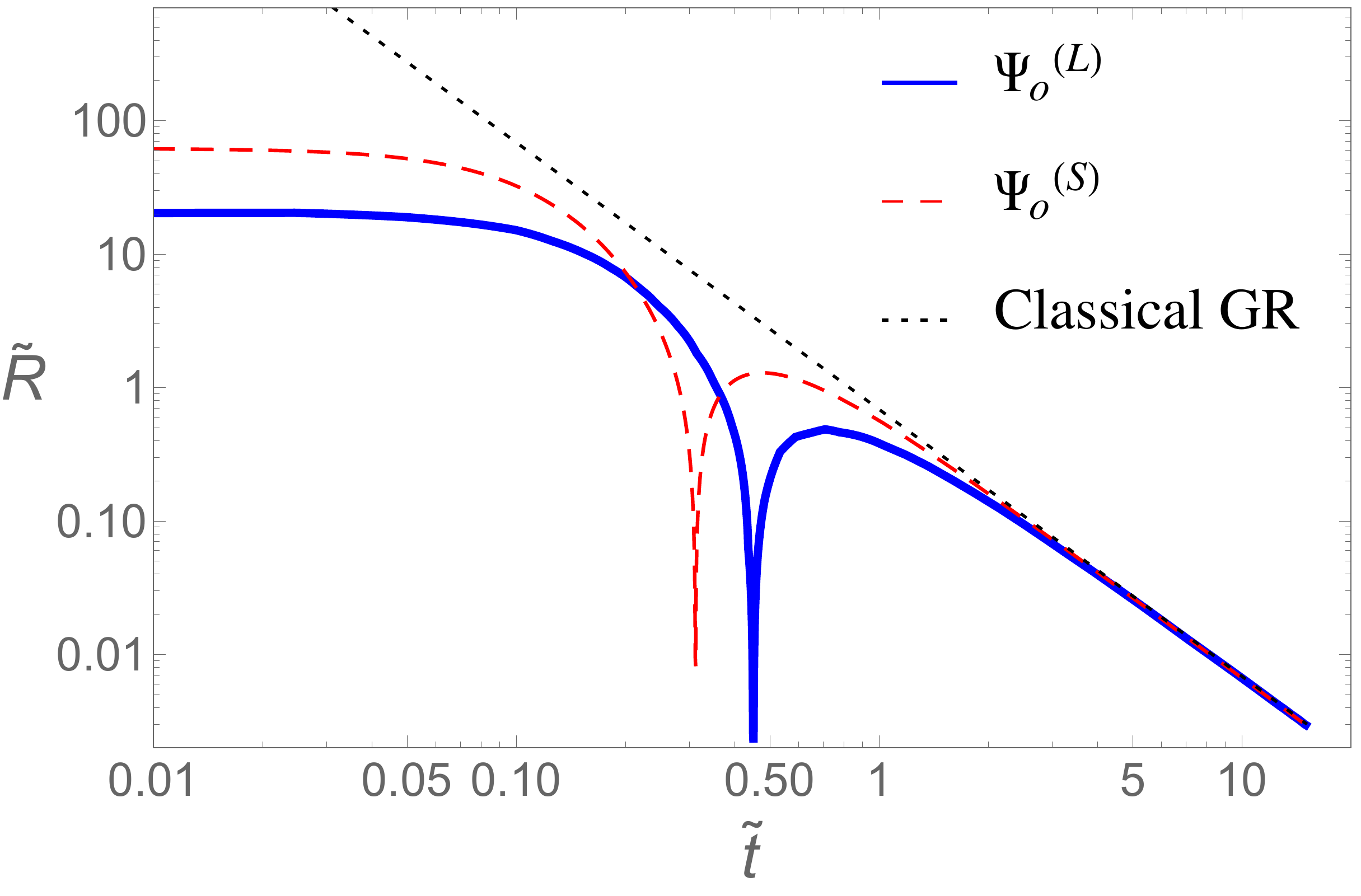}
\caption{Scalar curvature $\t{R}$ of the dressed effective metric $\t{g}_{ab}$ as a function of the cosmic time $\t{t}$ of $\t{g}_{ab}$. The continuous (blue) line shows the plot for a widely spread Gaussian $\Psi_{o}^{(L)}$ with large (i.e., $100\%$) relative dispersion $\Delta V/\b{V}$ in volume at the bounce. This evolution of $\t{R}$ is contrasted with that in a state $\Psi_{o}^{(S)}$ that is sharply peaked at the bounce with a small (i.e., $1\%$) relative dispersion represented by the (red) dashed line. At the bounce, the scalar curvature in the dressed effective geometry of  $\Psi_{o}^{(L)}$ is smaller than that in the sharply peaked state $\Psi_{o}^{(S)}$, but just $\sim 5 \spl$ after the bounce the two plots become indistinguishable, and furthermore, they coincide with the evolution of $\t{R}$ in classical general relativity, shown in dotted (black) line. This is why it is sufficient to compute the dressed effective metric only for $\sim 10 \spl$ after the bounce. Note that in general relativity $\t{R}$ grows unboundedly at earlier times because of he big bang singularity. The dip in $\t{R}$ in LQC --which occurs because there is a short period in which the equation of state of the inflaton approximates that of a radiation field \cite{aan3}-- is irrelevant for our discussion.}
\label{fig:dressed}
\efig

\subsection{The state $\Psi_{o}$ in the Planck era}
\label{s3.1}

The main question we wish to address in this paper is whether large quantum fluctuations in 
the background FLRW geometry leave observational imprints on CMB. Previous investigations in LQC have shown that there is an unforeseen interplay between the ultraviolet and the infrared: quantum geometry effects in the ultraviolet that tame the big-bang singularity affect the dynamics of scalar and tensor perturbations only at the largest observable angular scales, or low angular multipoles $\ell$ values \cite{aan3,am,agulloassym,ag3}. At these low $\ell$, observational error bars in the PLANCK mission data are rather large,  $\gtrsim 5\%$. Therefore, to simplify the detailed analysis we will use only those approximations that introduce errors of less than $\sim 0.001\,\%$ in our theoretical analysis.

For classes of states we consider, full quantum evolution given in Eq. (\ref{eq:lqchamilt}) is needed only in the `Planck era' where the matter density and curvature are \emph{greater than} $\sim 10^{-4}$ in Planck units, because in the subsequent evolution the dressed metric is extremely well approximated by a solution to Einstein's equation. As Fig. \ref{fig:dressed} shows, in our case this occurs already after $\sim 5 \spl$ after the bounce. However, we will compute the LQC evolution up to $\sim 10 \spl$ or more after the bounce  to ensure that there are no surprises and the strategy is reliable. 

Next, a general argument \cite{aan3} implies that the pre-inflationary LQC evolution can leave observational imprints only if the number $\Nbstar$ of e-folds between the bounce and the onset of slow roll inflation%
\footnote{\label{1} In this paper, the onset of slow-roll is defined to be the time  when the pivot mode $k_\star=0.002~\mpc$ exits the Hubble radius.}
is less than $20$. (The main part of this reasoning is summarized in section \ref{s4}.)  Thus, if $\Nbstar > 20$, then we already know that there will be no observational imprints of the quantum gravity regime --and therefore, in particular, of the possible large fluctuations in $\Psi_{o}(v,\phi)$-- on CMB. Therefore we will consider only those solutions $\Psi_{o}(v,\phi)$ in which $\Nbstar \le 20$ for both the quadratic and Starobinsky potentials. This in turn implies that the kinetic energy of the inflaton dominates over the potential energy throughout the quantum gravity regime that follows the bounce. For concreteness, let us consider the quadratic potential. Then, in the generalized effective trajectory $(\gGE, \phi)$ defined by $\Psi_{o}(v,\phi)$, potential energy is less than $\sim 10^{-12}$ times the kinetic energy in the Planck era. Therefore, in the evolution equation (\ref{eq:lqchamilt}), we can regard $\Theta_{0}$ as the `main part' and $\Theta_{1}$ as a perturbation. As explained in section \ref{s2.1}, we can construct the Hilbert space $\Hp$ by setting $\Theta=\Theta_{0}$ in (\ref{eq:lqchamilt}), i.e., by retaining only the `principal part' of $\Theta$. We will now use the inner product on this $\Hp$ to estimate how large an effect the potential $V(\phi)$ has on the evolution of $\Psi(v,\phi)$ during the Planck epoch.

Let us denote by $\Psi_{o}^{(0)}$ the states in $\Hp$, where the superscript $(0)$ is a reminder that we have set $\Theta=\Theta_{0}$. In $\Hp$, the second order equation (\ref{eq:lqchamilt}) is reduced to the `positive frequency' first order equation (\ref{evo}). To solve this equation we need to specify only the wave function $\Psi_{o}^{(0)} (v,\phi )$ at $\phi=\phib$. We take it to be a widely spread Gaussian with a $57\%$ relative dispersion in volume at $\phi=\phib$. This is smaller than the $100\%$ relative dispersion used in the rest of the paper because of certain numerical difficulties arise while solving the second order evolution equation (\ref{eq:lqchamilt}).
These are summarized at the end of this sub-section. However, for simulations we could perform within numerical limitations, results for the widely spread non-Gaussian state, and for the Starobinsky potential are essentially the same. Therefore we are confident that results reported in this sub-section will not change qualitatively if we had used additional resources and machine-time that are necessary to evolve states with $100\%$ relative dispersion.

\bfig
\ig[width=0.7\textwidth]{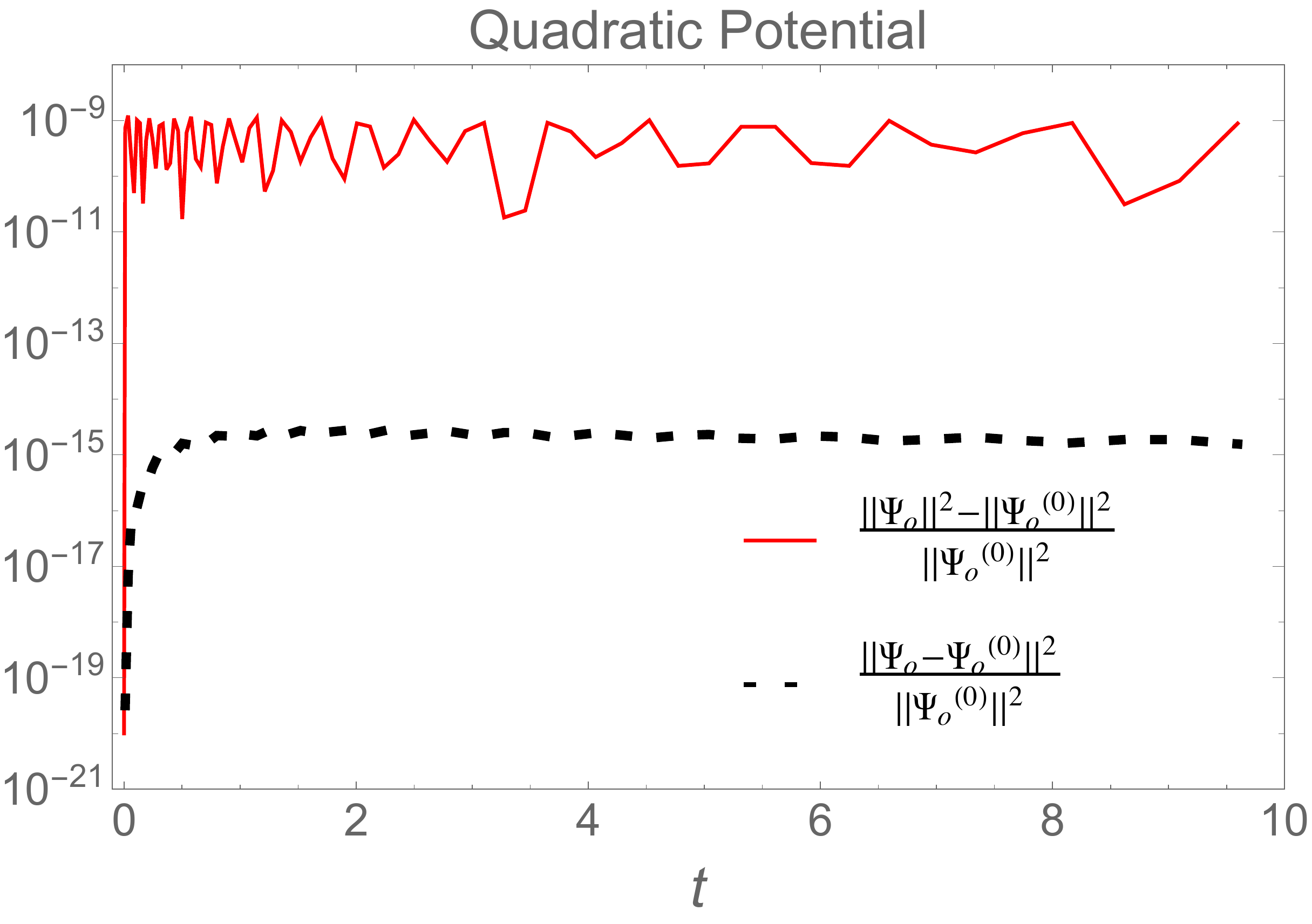}
\caption{Error involved in approximating $\Psi_{o}$ by $\Psi_{o}^{(0)}$. The two curves show that the error is less than one part in $10^{9}$, whence the approximation is excellent for our purposes.}
\label{fig:diff}
\efig

To solve the full second order equation (\ref{eq:lqchamilt}) with the potential $V(\phi)$, we need to specify both the wave function $\Psi_{o}(v,\phib)$ and its first derivative $\partial_{\phi}\Psi_{o}(v,\phi)\mid_{\phi=\phib}$. Since we wish to compare the solution $\Psi_{o}$ to the second order equation with $\Psi_{o}^{(0)}$, the initial data at $\phi=\phib$ for $\Psi_{o}$ will be constructed from $\Psi_{o}^{(0)}$. That is, at $\phi = \phib$ we will set $\Psi_{o}$ and its first $\phi$-derivative equal to $\Psi_{o}^{(0)}$ and its first $\phi$-derivative. We then evolve $\Psi_{o}$ using the second order equation (\ref{eq:lqchamilt}) across the Planck regime. We use two quantities to measure the difference between the solution $\Psi_{o}$ to the full, second order equation, and the  solution $\Psi_{o}^{(0)}$ to the equation in which the potential $V(\phi)$ --and hence $\Theta_{1}$-- is set to zero:
\begin{enumerate}
\item The relative difference  
\be I(\t{t}) := \f{||\Psi_{o}||^{2}\, -\, ||\Psi_{o}^{(0)}||^{2}}{||\Psi_{o}^{(0)}||^{2}}\,\, (\t{t}) \ee 
as a function of cosmic time $\t{t}$,\, defined by the dressed metric $\t{g}_{ab}$ of $\Psi_{o}^{(0)}$. Since, as explained in section \ref{s2.1}, the norm of $\Psi_{o}^{(0)}$ is conserved during the evolution, this difference measures the error involved in using the norm in $\Hp$ for the wave function $\Psi_{o}$ that satisfies the full (second-order) evolution equation (\ref{eq:lqchamilt}), i.e. the amount by which the norm on $\Hp$ fails to be conserved by the full evolution. (We can of course just use normalized states so that the denominator is $1$.) 

\item The relative difference 
\be II(\t{t}):= \f{||\Psi_{o}\, - \, \Psi_{o}^{(0)}||^{2}}{||\Psi_{o}^{(0)}||^{2}}\,\, (\t{t}) \ee
again, as a function of cosmic time $\t{t}$,\, defined by the dressed metric $\t{g}_{ab}$ of $\Psi_{o}^{(0)}$. This is a direct measure of the error one makes in replacing $\Psi_{o}$ by $\Psi_{o}^{(0)}$ at time $\t{t}$. (Again, we can of course just use normalized states so that the denominator is $1$.)
\end{enumerate}

Results are shown in Fig. \ref{fig:diff}. The solid (red)  curve shows the quantity $I(\t{t})$ and the dashed (black) curve shows $II(\t{t})$. Since $\t{t}=0$ represents the bounce time and since we have used the same initial data for the two states there, both quantities vanish at this initial time. They grow immediately after the bounce but quickly plateau. Throughout the Planck regime $I \lesssim 10^{-9}$ and $II < 10^{-14}$. The error in $II$ is less than that in $I$ because of the interference terms. In either case we see that the approximation is excellent since the error involved is much less than our tolerance level $ 0.001\%$. For the Starobinsky potential, along the generalized effective trajectory the ratio of the potential energy to kinetic remains less than $10^{-8}$ throughout the Planck regime (for quadratic this figure was $10^{-12}$). In this case, $I(\t{t})$ is bounded by $10^{-7}$ and $II(\t{t})$ is bounded by $10^{-11}$. While these values are higher than those for the quadratic potential, they are nonetheless well below our tolerance level. 

To summarize, then, because the kinetic energy dominates over the potential energy at the bounce, throughout the Planck regime where we need the full LQC evolution, the wave function $\Psi_{o}$ to the full Hamiltonian constraint is approximated \emph{extremely well} by the solution $\Psi_{o}^{(0)}$ to the Hamiltonian constraint where the term $\Theta_{1} \Psi_{o}$ is simply dropped. Therefore, in the remainder of the paper we will
simply use the first order `positive frequency' evolution equation (\ref{evo}). Then, as remarked in section \ref{s2.1}, the evolution can be carried out by calculating once and for all the eigenfunctions $e_{k}(v)$ of the operator $\Theta_{0}$: The solution to (\ref{evo}) is then
\be \label{state2} \Psi_{o}^{(0)}(v,\phi) = \int_{-\infty}^{\infty}\!\! \dd\underbar k \,\,\t{\Psi}(\underbar k)\,\, e_{\underbar k}(v)\, e^{i\omega\phi}  \quad {\rm with}\quad \omega(\underbar{k}) = \sqrt{12\pi G} |\underbar k|\, ,\ee
where $\t{\Psi}(\underbar k)$ is the profile function that determines the relative dispersions in the state $\Psi_{o}^{(0)}$ at the bounce. This method avoids the second order equation altogether but is not directly available if one uses $\Theta$ in place of $\Theta_{o}$ because of $\phi$-dependence and self-adjointness issues associated with full operator $\Theta$ \cite{ap}. \emph{From now on, for simplicity of notation we will drop the superscript $(0)$ and denote this solution simply by $\Psi_{o}(v,\phi)$.}  \\

We will conclude this discussion with a brief summary of the numerical difficulties associated with the second order equation (\ref{eq:lqchamilt}), for states which have relative dispersions greater than $\sim 57\%$ at the bounce. Note first that because the kinetic energy dominates over the potential energy throughout the Planck regime, we have $V(\phi) \ll  \omega^{2}(\underbar k_{o})/ 2\bar{V}^{2}$\,\, where $\underbar k_{o}$ is the value at which the profile function $\t\Psi(\underbar k)$ in (\ref{state2}) is sharply peaked and $\bar{V}$ is the expectation value of $\hat{V}|_{\phi}$. However, since the operator $\Theta_{1}$ is given by $\Theta_{1} = {\rm const}\,\,\, V(\phi) v^{2}$, for sufficiently large values of $v^{2}$, the product  $V(\phi) v^{2}$ can become large. Nonetheless, because the wave function $\Psi_{o}(v,\phi)$ goes to zero \emph{very} rapidly for large $v$,\,\, $\Theta_{1}\Psi_{o}$ is still very small for these large $v$. But because we are using a second order scheme with double precision in our numerics, the code does not distinguish between values  $\sim 10^{-18}$ of wave functions and those which are \emph{much} smaller. Therefore, the numerical calculation overestimates the value of $\Theta_{1}\, \Psi_{o}(v,\phi)$ for sufficiently large $v$, giving a larger answer than the correct value. Larger the relative dispersion in $v$, larger is the domain in $v$ we need to consider, and at some stage the second order scheme with double precision ceases to be adequate. It turned out that we could run the code based on double precision, developed in \cite{dgs1}, accurately if the relative spread is $57\%$ at the bounce. But to go beyond, we would need a higher precision scheme in addition to large domain in $v$ which would require considerable new numerical work and the runs would also be significantly more expensive computationally. For example, the numerical evolution of a wavefunction with the relative spread $57\%$ at the bounce (shown in \fref{fig:diff}) using the second order scheme with double precision requires $\sim10$ hours on a modern computer with 16 cores. The same numerical evolution with quadruple precision will take at least ten times more computing time. That is why in this sub-section we restricted ourselves to states with relative dispersion up to $57\%$. We believe that the final results will be the same if one has an appropriately modified code that can handle larger dispersions.

\subsection{Dressing of the scale factor and conformal time}
\label{s3.2}

Given $\Psi_{o}(v,\phi)$, we can calculate the dressed effective metric $\t{g}_{ab}$. It is determined by the dressed scale factor $\t{a}$ and the dressed conformal time $\t\eta$ defined in Eqs. (\ref{dresseda}) and (\ref{dressedeta}). The scale factor operator $\h{a}$ that features in these expressions is given in expression (\ref{ahat}). Since the operator $\h{H}_{o}$ that also features in the expression is given by $\hat H_{o} = \sqrt{\Theta}_{0}$ (see Eq. (\ref{evo})), its action is simplest in the eigenbasis of the $\Theta_{0}$ operator (see Eq. (\ref{state2})). Thus,
\be
 \hat H_0 \, \Psi_o(v,\phi) = \hbar \int d\underbar k\,\omega\,
 \t \Psi(\underbar k)\, e_{\underbar k}(v) \, e^{i\omega\phi}\, , \quad {\hbox{\rm where, as before}} \quad \omega({ \underbar k}) = \sqrt{12\pi G} |\underbar k|\, .
\ee
Therefore, once the frequency content $\tilde \Psi(\underbar k)$ of the state $\Psi_o(v,\phi)$ is specified, we have
\be
 \tilde a(\phi) = \Big(\f{1}{V_{o}}\Big)^{\f{1}{3}} \,\,\left(\f{\langle \int d\underbar k\,\omega^{-1/2}\, \tilde\Psi(\underbar k)\, 
e_{\underbar k}(v) \, e^{i\omega \phi}\, |\, \int d\underbar k\,\omega^{-1/2}\, \tilde\Psi(\underbar k)\, 
V^{4/3}e_{\underbar k}(v) \, e^{i\omega \phi}\rangle}
{\langle \int d\underbar k\,\omega^{-1/2}\, \tilde\Psi(\underbar k)\,
e_{\underbar k}(v)\,  e^{i\omega \phi}\, |\, \int d\underbar k\,\omega^{-1/2}\, \tilde\Psi(\underbar k)\,
e_{\underbar k}(v) \, e^{i\omega\phi}\rangle}\right)^{1/4}.
\ee
where $V=2\pi G v$, and for reasons explained in section \ref{s3.1}, we use the inner product determined by the norm of Eq.(\ref{norm}):
\be\label{ip}
\langle \Psi_{o}| \chi_{o}\rangle\, :=\, \sum_{v=4n\ell_{o}} \bar{\Psi}_{o} (v,\phi)\, \chi_{o}(v,\phi) \, .\ee 
Similarly, the dressed conformal time is given by 
\be
 \dd\tilde \eta = V_o~\tilde a^2~\langle H_0^{-1} \rangle~ \dd\phi\, .
\ee

\bfig
 \ig[width=0.45\textwidth]{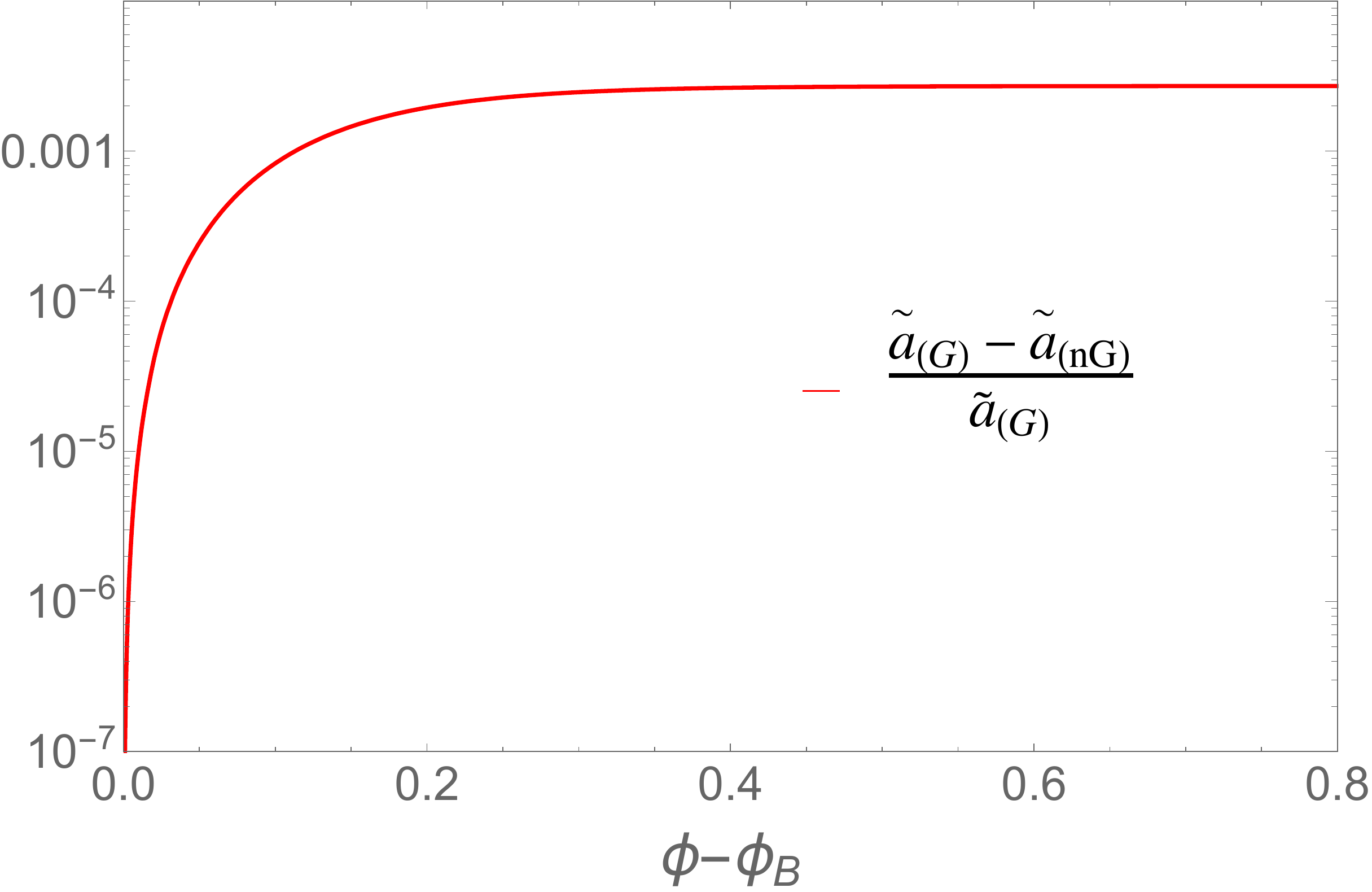}
\hskip0.2cm
 \ig[width=0.45\textwidth]{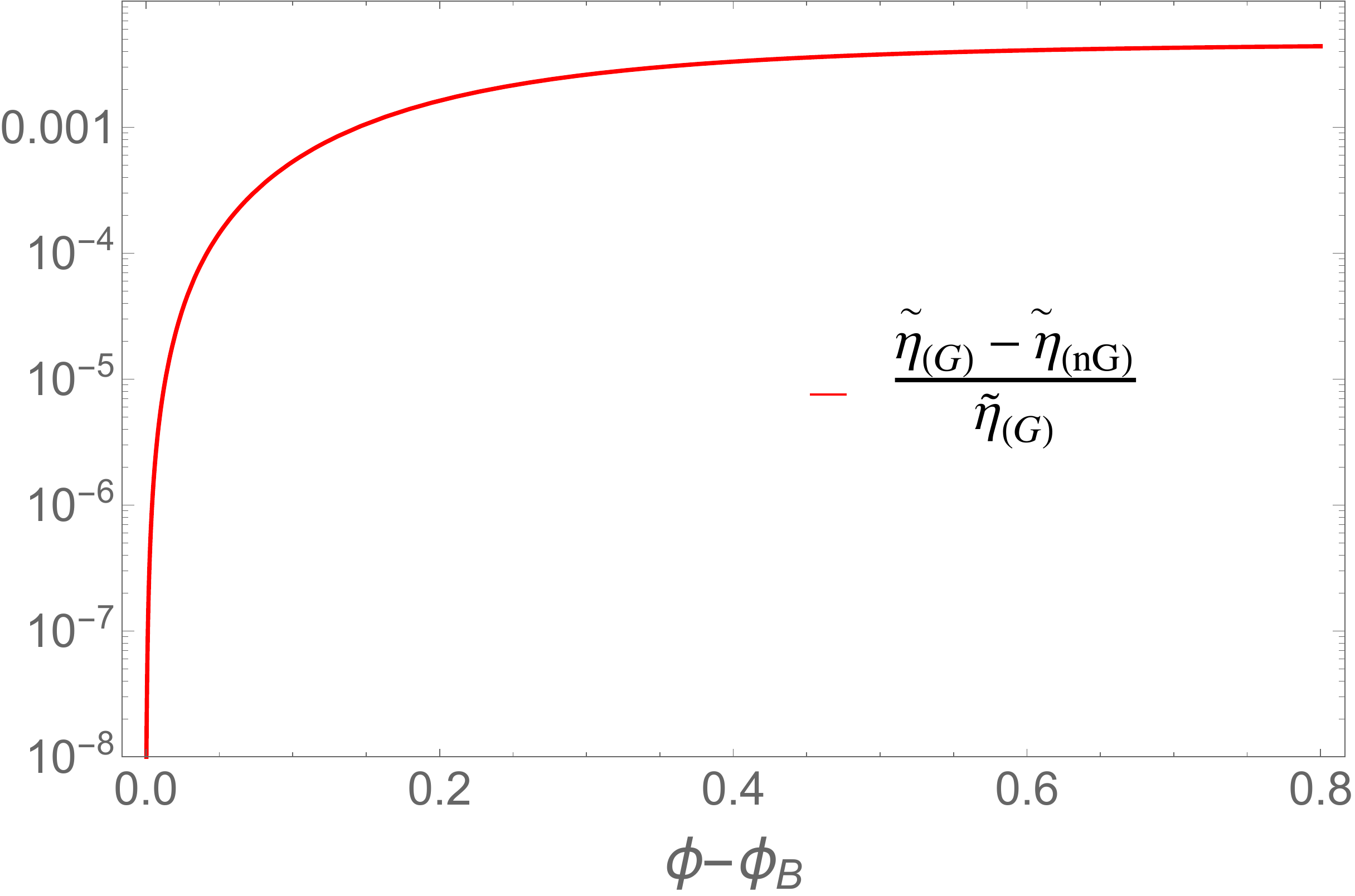}
\caption{Comparison between the dressed effective geometries determined by Gaussian and non-Gaussian widely spread states with $100\%$ relative dispersions in the volume at the bounce. \emph{Left Panel:} Comparison between the dressed effective scale factors $\t{a}$. \emph{Right Panel:} Comparison between the dressed effective conformal times $\t\eta$. Even though the states have very different profiles and therefore quantum fluctuations, the relative differences are $\lesssim 0.1\%$.}
\label{fig:dressed-comparison}
\efig

\bfig
 \ig[width=0.45\textwidth]{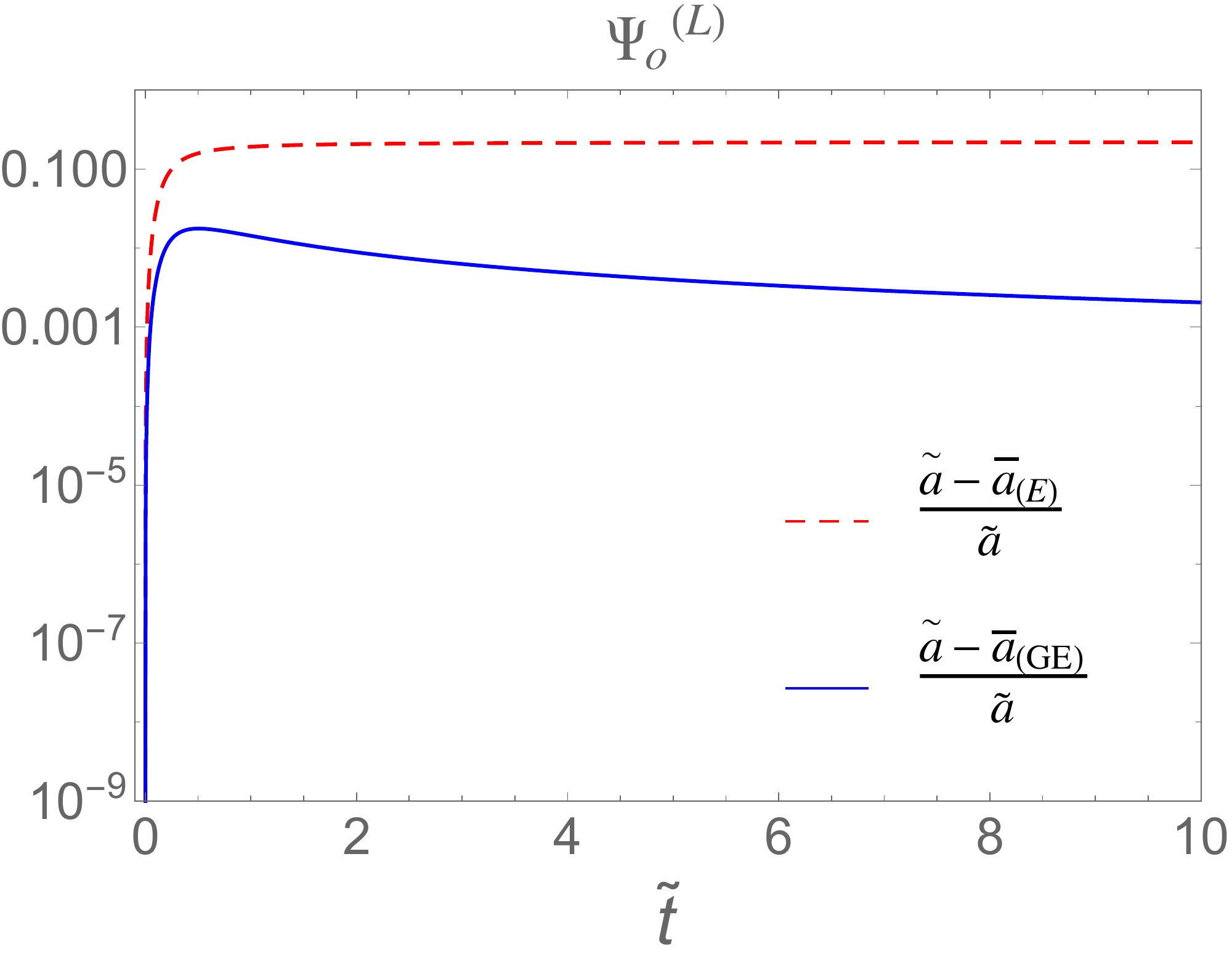}
\hskip0.2cm
 \ig[width=0.45\textwidth]{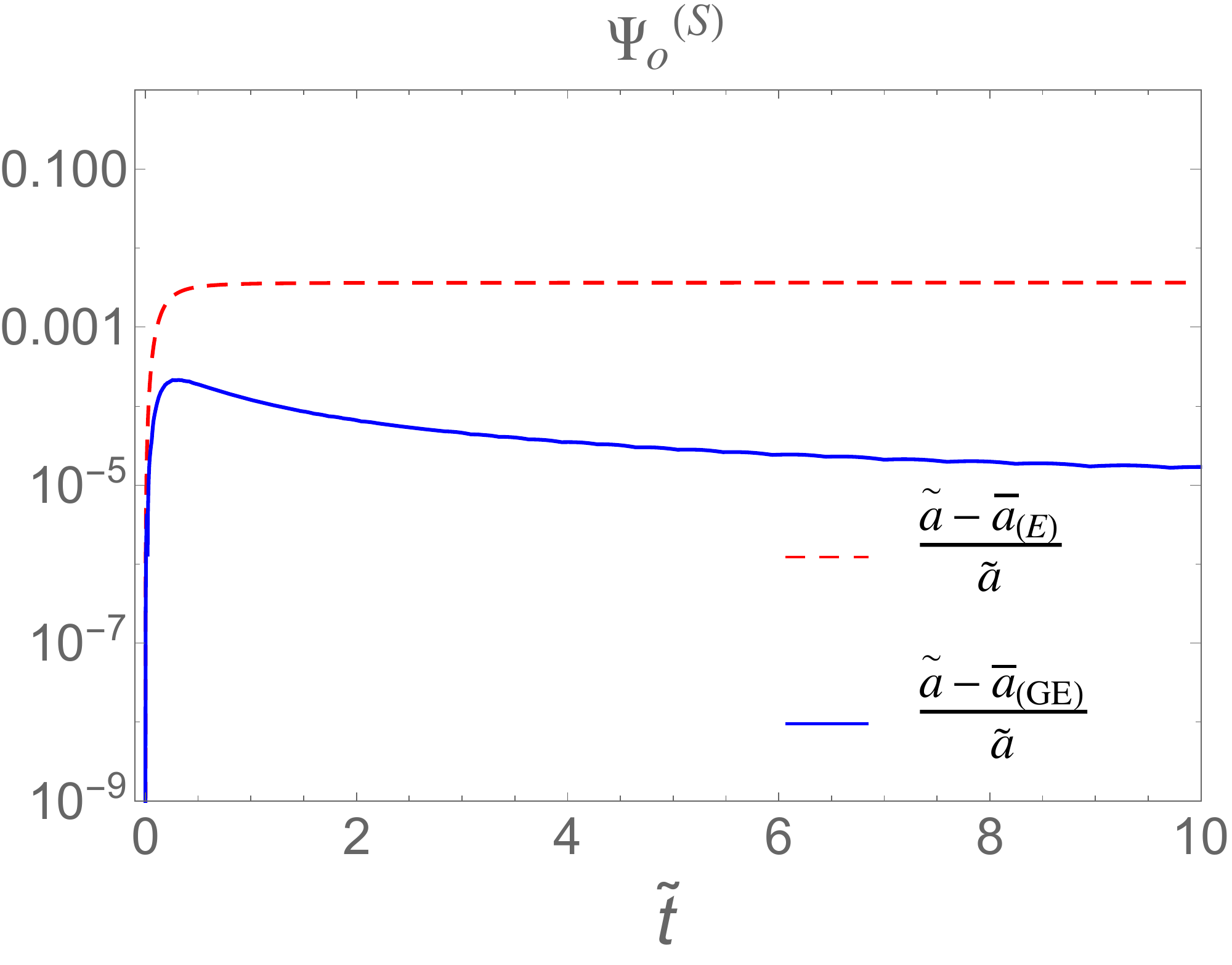}
\caption{Comparison between the time evolution of three scale factors: $\t{a}$ of the dressed effective metric $\t{g}_{ab}$, $\bar{a}_{GE}$ of the generalized effective metric $\b{g}_{ab}^{\rm (GE)}$, and  $\bar{a}_{E}$ of the generalized effective metric $\b{g}_{ab}^{(E)}$. \emph{Left Panel:} Plots of the Gaussian state $\Psi_{o}^{(L)}$ with large ($100\%$) relative dispersions in the geometry at the bounce. \emph{Right Panel:} Plots of the sharply peaked Gaussian state $\Psi_{o}^{(S)}$ with small ($1\%$) relative dispersions in the geometry at the bounce. In both cases, $\bar{a}_{\rm (GE)}$ approximates $\t{a}$ reasonably well, to an accuracy of $\sim 0.1\%$ for $\Psi_{o}^{(L)}$ and $0.01\%$ for $\Psi_{o}^{(S)}$. But $\bar{a}_{\rm (E)}$ \emph{does not} approximate $\t{a}$ well; for $\Psi_{o}^{(L)}$ the discrepancy is $\sim 10\%$.}
\label{fig:dressedComparison}
\efig

The sums and integrals in the above expressions are computed numerically. Although the summation over $v$ has to be performed over the entire lattice, for numerical computation one has to restrict to a finite domain. In our computation we take the finite domain to be $[0,\,v_{\rm max}]$, where $v_{\rm max}$  is chosen so that the amplitude of the wave function $\Psi_o$ for $v\ge v_{\rm max}$ is less than $10^{-10}$ times its peak value. The numerical computation of the dressed effective metric of a typical state with $\Delta V/V\approx 1$ we used  takes approximately 3 hours on a modern single node computer with 20 parallel cores. Furthermore, evolution of scalar modes on this dressed metric and the computation of the power spectrum at the end of inflation, discussed in section \ref{s4}, requires additional 3 to 4 hours on the same computing node. Thus, in total, each simulations takes approximately 7 hours. In order to perform large number of simulations, we used modern high performance computing system (HPC) with multiple nodes, where  individual simulations run on different nodes.

In the two panels of Fig. \ref{fig:dressed-comparison}, we compare the dressed scale factor $\t{a}$ and the dressed conformal time $\t\eta$ for the Gaussian and non-Gaussian states with $100\%$ dispersion in the volume at the bounce. Although in these two states the expectation value and uncertainty in volume is the same at the bounce, as Fig. \ref{fig:wavefun} shows, their profiles are \emph{very} different. Hence, the higher `moments' --i.e., the expectation values and dispersions in higher powers of the volume operator-- are also quite different already at the bounce. Yet, throughout the relevant time interval starting from the bounce, the relative differences in $\t{a}$ and $\t\eta$ are $\lesssim 0.1\%$. (Outside this interval, as Fig. \ref{fig:dressed} shows, the dressed metric is essentially indistinguishable from the FLRW solution to classical Einstein's equation.) Note that the detailed quantum geometries determined by the Gaussian and the non-Gaussian states are very different. If one were to go and measure higher order moments in say matter density or curvature in them one would find very different values. But, as discussed in section \ref{s2.3}, cosmological perturbations are blind to these differences; their dynamics is governed only by the dressed metric $\t{g}_{ab}$ determined by the two wave functions. And because two dressed metrics  are so close to one another, cosmological perturbations do not distinguish them.

To gain intuition about the effects of the dispersion of the state $\Psi_o$, in \fref{fig:dressedComparison} we compare the evolution of the dressed scale factor $\t{a}$ with that of the scale factor $\aGE$ obtained by solving the generalized effective equations (\ref{GEE}), and the scale factor $\aE$ obtained by solving the effective equations (\ref{EE}). The left panel shows this comparison for widely spread Gaussian state (with $100\%$ relative dispersion in volume at the bounce) and the right panel for the sharply peaked Gaussian (with $1\%$ relative dispersion). These plots serve to bring out two features: (i) While effective equations (\ref{EE}) approximate the dynamics of dressed scale factor reasonably well (to $0.1\%$ accuracy) for sharply peaked states, \emph{they do a poor job for the widely spread state} (where the relative difference is $10\%$). This point is important to bear in mind because effective equations are very widely used in LQC; and, (ii) Generalized effective equations provide a significantly better approximation even for sharply peaked states (accuracy of 1 part in $10^{5}$ versus $10^{3}$ for effective equations). Thus, even for states which have only $1\%$ relative dispersion in volume, the difference between $\rhob$ and $\rhosup$ that distinguishes (\ref{GEE}) and (\ref{EE}) can not be ignored if the goal is to have high accuracy in dynamics of cosmological perturbations.

\section{Fluctuating FLRW backgrounds and Observations} 
\label{s4}

This section is divided into two parts. In the first we compute of the power spectrum $\mathcal{P}_{\mathcal{R}}(k)$ and the spectral index $n_{S}(k)$ for scalar perturbations propagating on Gaussian and non-Gaussian FLRW quantum geometries $\Psi_{o}(v,\phi)$ with large relative dispersions. As a check on robustness we consider Starobinsky as well as quadratic potentials. The main findings are: (i) the qualitative behavior of  $\mathcal{P}_{\mathcal{R}}(k)$ for states $\Psi_{o}(v,\phi)$ with large fluctuations in geometry is similar to that of sharply peaked states; and, (ii) $\mathcal{P}_{\mathcal{R}}(k)$ and $n_{S}(k)$ for the two classes of widely spread states are \emph{essentially indistinguishable}, even though the non-Gaussian state has  \emph{very} different profile from the Gaussian state. In the second part, we report an even more surprising finding: within observational errors, $\mathcal{P}_{\mathcal{R}}(k)$ and $n_{S}(k)$ that result from widely spread states are \emph{indistinguishable} from those that result from sharply peaked states after a small adjustment in the number $\Nbstar$ of pre-inflationary e-folds for the sharply peaked states. 

\subsection{Computation of the power spectrum}
\label{s4.1}

We will now use the dressed effective metrics calculated in section \ref{s3.1} and the evolution equations for scalar modes discussed in section \ref{s2.3} to calculate the power spectrum. We follow the same strategy as in previous LQC analyses \cite{aan1,aan3,am,bg1,bg2}: starting from initial conditions for perturbation near the bounce, we evolve them through the quantum gravity regime until the end of the subsequent inflationary phase, and compute power spectra at that time. Since our focus is on the observable effects, we will  discuss only the scalar modes. The steps involved in the analysis of tensor modes are identical and our conclusions are the same as those for the scalar modes.

Quantum theory of the Mukhanov-Sasaki scalar perturbations $\h{\Q}(\vec{x},\t\eta)$ follows standard steps (see e.g. \cite{aan1} for a summary). The homogeneity of the dressed metric $\tilde g_{ab}$ makes it convenient to perform a Fourier expansion of $\h{\Q}(\vec{x},\t\eta)$:
\be
  \h{\Q}({\vec x},\etat) = \int\f{{\rm d}^3 k}{(2\pi)^3} \,  \h{\Q}_{\vec{k}}( \etat)\, e^{i{\vec k}\cdot{\vec x}}= \int\f{{\rm d}^3 k}{(2\pi)^3} \left(\hat A_{\vec k}~q_k(\etat) + \hat A^\dagger_{-\vec k}
~q_{k}^*(\etat)\right) e^{i{\vec k}\cdot{\vec x}},
\ee
where in the second equality we have written  $\h{\Q}_{\vec{k}}( \etat)$ in terms of the creation and annihilation operators $\hat A_{\vec k}$ and $\hat A^\dagger_{\vec k}$, and a basis of mode functions $q_{k}(\etat)$. These basis functions satisfy the equation motion
\be
 q_k^{\prime\prime} (\etat) + 2\f{ \at'(\etat)}{\at(\etat)} q_k^\prime(\etat) + (k^2 +
\tilde{\u }(\etat))\,  q_k(\etat) =0\, ,
\ee
and the normalization condition
\be  q_k(\etat) q_k'^*(\etat)-q^*_k(\etat) q'_k(\etat) = \frac{i}{\at(\etat)^2} \, ,\ee
where  $k=|{\vec k}|$. As usual, the scalar power spectrum of $\mathcal P_{\!\mathcal{Q}}$, is extracted from the two-point function in momentum space via
\be
  \langle 0|\h{\Q}_{\vec k}(\etat) \h{\Q}_{\vec k^\prime}(\etat)|0\rangle =:
(2\pi)^3\delta({\vec k}+{\vec k^\prime}) \f{2\pi^2}{k^3} \mathcal P_{\!\mathcal
Q}(k, \etat)\, ,
\ee
where $|0\rangle$ is the vacuum annihilated by the operators $\hat A_{\vec{k}}$ for all $\vec{k}$. In terms of mode functions, we have  $\mathcal P_{\!\mathcal{Q}}(k,\eta) = (\hbar\,{k^3}/{2\pi^2})\,|q_k(\eta)|^2$. The power spectrum of comoving curvature perturbations at the end of inflation, is then obtained as 
\be \mathcal  P_{\mathcal{R}}(k): =
\bigg(\frac{H(\etat_{\rm end})}{\dot\phi(\etat_{\rm end})}\bigg)^2\mathcal  P_{\!\mathcal{Q}}(k,
\etat_{\rm end})\, .\ee 
In order to compute this power spectrum in LQC we need to carry out following steps:

\begin{enumerate} 

\item Specify a potential $V(\phi)$ for the inflaton field. As discussed in the last section, we will use the quadratic and Starobinsky potentials, 
\be V(\phi)=\frac{1}{2}m^2\, \phi^2, \quad {\rm and}\quad V(\phi)=\f{3m^2}{32 \pi G}\,\, \Big(1-e^{-\sqrt{\f{16\pi G}{3}}\phi}\Big)^2 ,\quad {\rm respectively} \, , \ee 
where, in each case, the value of $m$ is derived from the PLANCK mission data  \cite{planck}. 

\item Specify a state for the homogeneous and isotropic background spacetime $\Psi_o(v,\phi)$. As discussed in section \ref{s2},  we consider Gaussian as well as multi-peaked non-Gaussian states $\Psi_o(v,\phi)$ with $100\%$ relative dispersion in volume at the bounce (see \fref{fig:wavefun}).

As discussed in section \ref{s3}, we assume that the kinetic energy of the inflaton $\phi$ dominates over the potential energy in the quantum gravity regime. In the subsequent evolution, the ratio of the potential energy to the total energy of $\phi$ grows and, at the time $t_{\star}$ when slow roll inflation begins, the potential energy dominates over kinetic. The number $\Nbstar=\ln \big(a(t_\star)/a(t_{\rm B})\big)$ of e-folds of between the bounce and $t_{\star}$ depends on the choice of $\Psi_{o}(v,\phi)$, and is a new parameter originating in LQC. For the two potentials under consideration, it can be traded with the value $\phib$ of the inflation at the bounce time. Since it is simpler to specify $\phib$ in calculations, the new parameter was taken to be $\phib$ in many LQC analyses. While both parameters contain the same information, now the focus has shifted to $\Nbstar$ because it has a more direct physical interpretation \cite{ag3}. As noted in section \ref{s3.1}, we are interested in the initial conditions for which $\Nbstar\lesssim 20$. For larger values of $\Nbstar$, physical wavelengths of modes in the CMB that are affected during the evolution in the quantum gravity regime turn out to be significantly larger than the radius of the observable universe. In these cases, then, the \emph{observational} difference between sharply peaked and widely spread states of quantum geometry is simply washed out during pre-inflationary expansion. Thus, for the question of whether quantum fluctuations in the background geometry in the Planck era have observational consequences, only the case $\Nbstar\lesssim 20$ is interesting. It turns out that this condition in turn implies that, at the bounce, $\phi_{B}$ is small, i.e., the kinetic energy of the inflation dominates over its  potential energy at the bounce.

\item  Specify the quantum state for scalar perturbations at some initial time. Since our goal is to isolate the effects of different choice for the \emph{background} quantum geometry $\Psi_o(v,\phi)$, to `compare apples with apples' we need to choose the same quantum state of perturbations, as we vary $\Psi_{o}(\phi, v)$. For concreteness, we will work with the `preferred instantaneous vacuum'  state,%
\footnote{This state is defined as the unique vacuum  for which the adiabatically renormalized energy-momentum tensor vanishes at a given instant.} 
introduced in \cite{ana}, at $0.1$ Planck seconds before the bounce time. For general reasons discussed in \cite{am}, we expect that other `natural' choices of initial conditions will not change our final results significantly, and we have verified this by allowing one other choice.
\end{enumerate}

\bfig
 \ig[width=0.85\textwidth,height=3in]{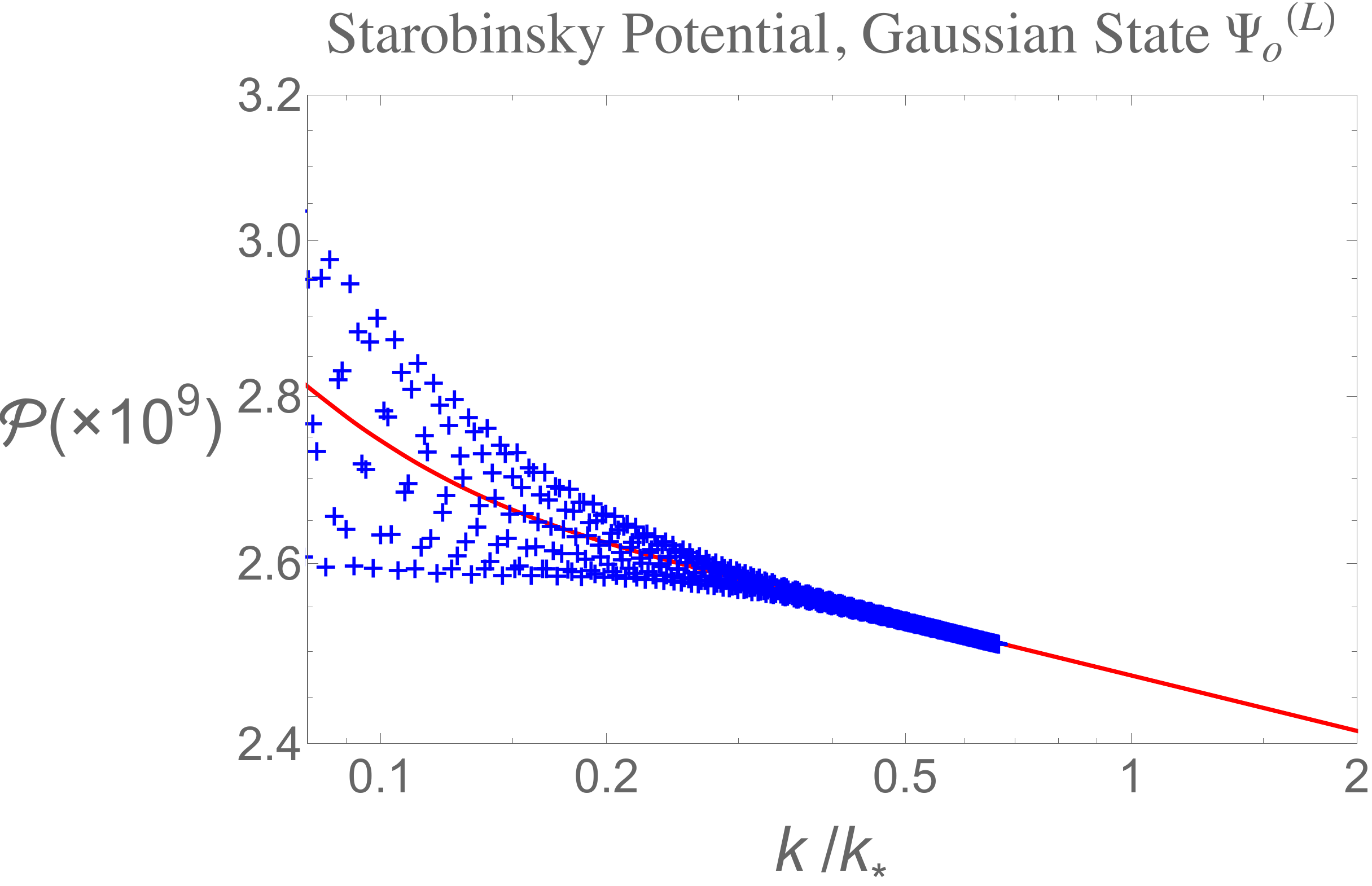}
\caption{Scalar power spectrum at the end of inflation for
Starobinsky potential for a widely spread Gaussian state with $100\%$ relative dispersion at the bounce and $\Nbstar=16.8$. Blue crosses are the numerical result for individual values of $k$. The red line results from averaging the high frequency oscillations with bins of size $0.25~\lp^{-1}$ in $k$ (with $\tilde{a}_{\rm B}=1$). (See \cite{aan3} for a discussion of these oscillations).} 
\label{fig:ppsns1}
\efig

In earlier LQC investigations \cite{aan3,am,agulloassym,ag3}, these steps were carried out for sharply peaked states. As explained in section 4.1 of \cite{aan3}, the pre-inflationary dynamics does leave imprint on the power spectrum, in that there are deviations from the standard inflationary predictions based on the use of the Bunch-Davies vacuum at $t=t_{\star}$. The scalar curvature $\t{R}_{\rm B}$ of the dressed metric $\t{g}_{ab}$ at the bounce introduces a new scale $k_{\,\rm LQC}$ given by $(k_{\,\rm LQC}/ \t{a}_{\rm B})^{2}= (\t{R}/6)$. For modes with $k\gg k_{\, \rm LQC}$ the effects of pre-inflationary dynamics are negligible. On the other hand modes with $k\lesssim 10\, k_{\, \rm LQC}$ are significantly affected during the evolution in the Planck regime. This effect translates into a departure from the power spectrum predicted by standard inflation for $k\lesssim 10 k_{\, \rm LQC}$.%
\footnote{Depending on the choice of initial conditions for perturbations, the power may be enhanced \cite{aan3} or suppressed \cite{ag3} at for $k \lesssim 10 k_{\, \rm LQC}$.}
Thus, there is an unforeseen interplay between the ultraviolet and the infrared: Quantum geometry effects that tame the singularity create a new scale $k_{\, \rm LQC}$ and it is the \emph{long wave-length} modes of cosmological perturbations with $k \lesssim 10 k_{\, \rm LQC}$ that are excited during the quantum gravity regime. 

\bfig
\ig[width=0.44\textwidth]{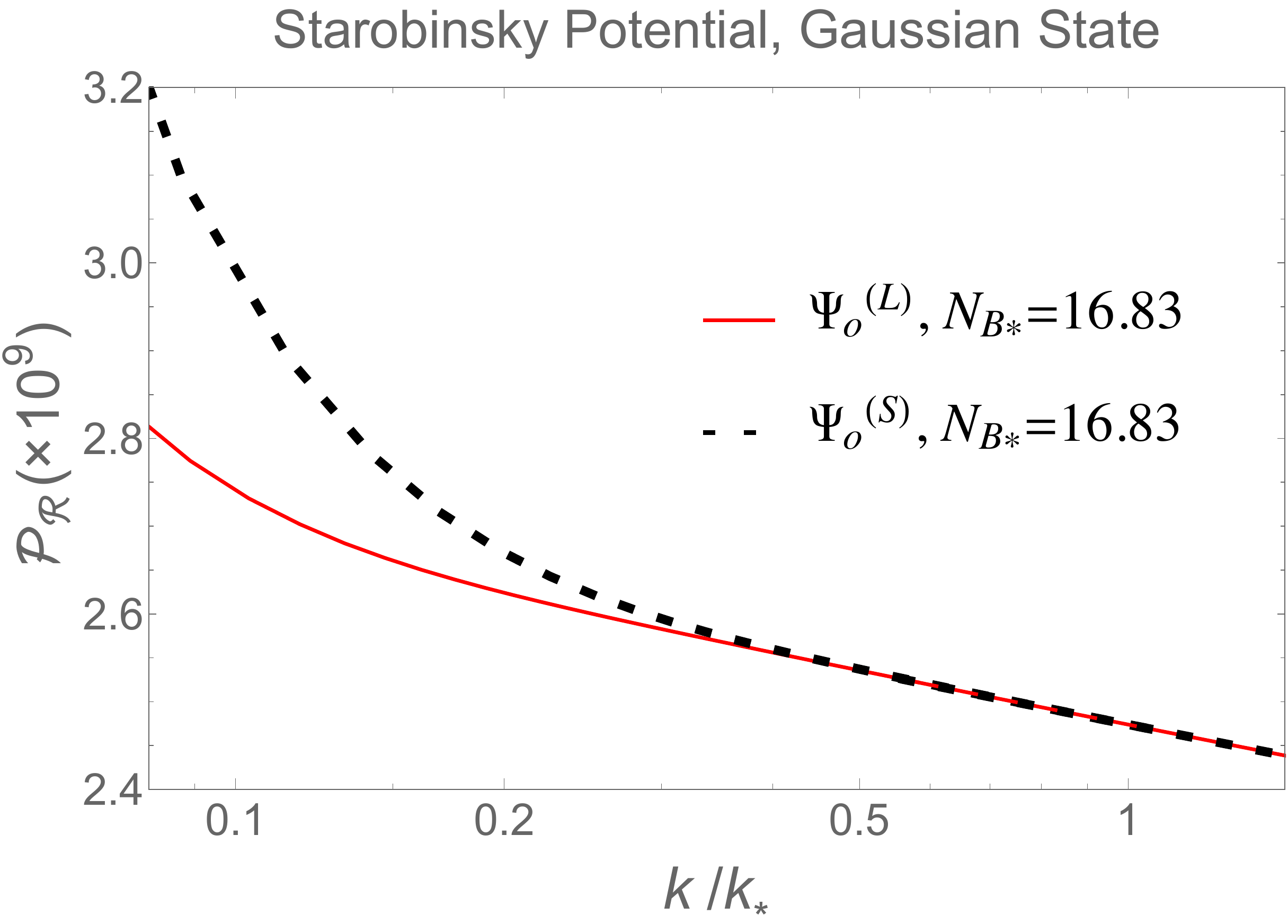}
\hskip0.3cm
\ig[width=0.44\textwidth]{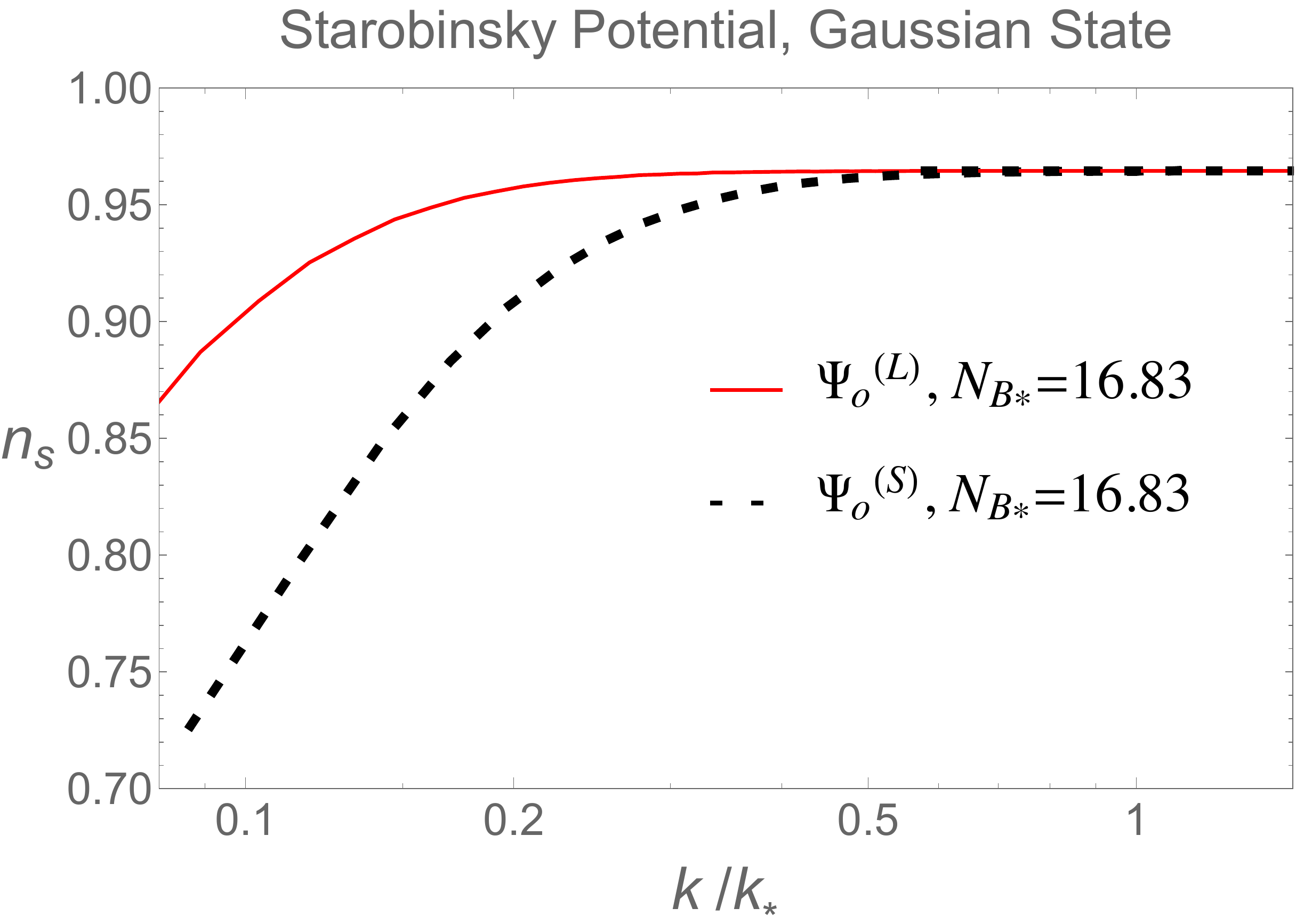}
\caption{{\it Starobinsky potential:} Comparison of observable predictions from a sharply peaked state $\Psi_{o}^{(S)}$ with $\Delta V/V=0.01$ at the bounce and a widely spread state with $\Delta V/V=1$ at the bounce. Both states are Gaussian.\,\,\,{\it Left panel:} Scalar power spectrum. {\it Right panel:} Scalar spectral index. 
Note that in both plots, the two curves appear to agree if the one resulting from sharply peaked states is shifted slightly to the left. This expectation is borne out by the detailed simulations reported in section \ref{s4.2}.}
\label{fig:powernsGaussStarobinsky}
\efig
\bfig
\ig[width=0.44\textwidth]{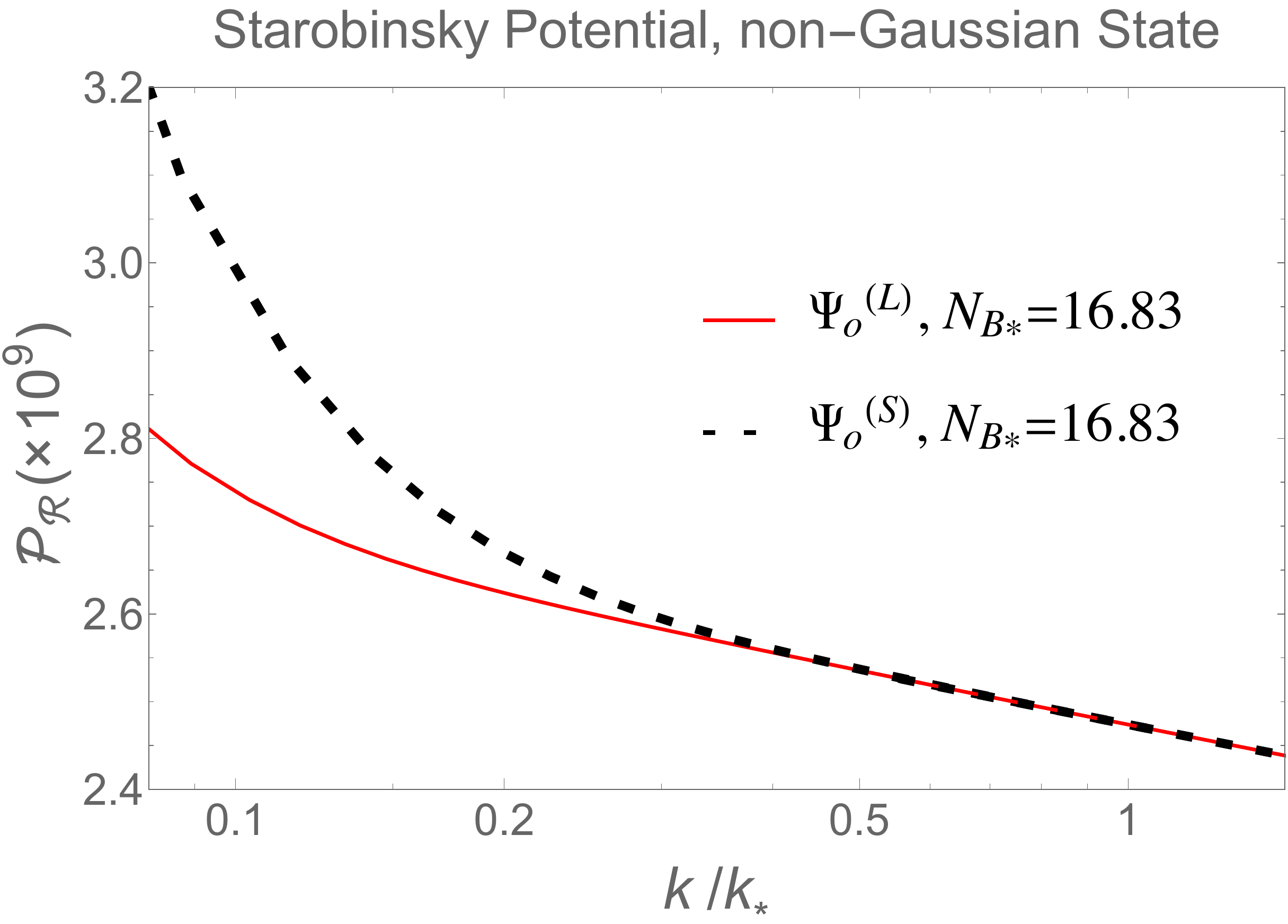}
\hskip0.3cm
\ig[width=0.44\textwidth]{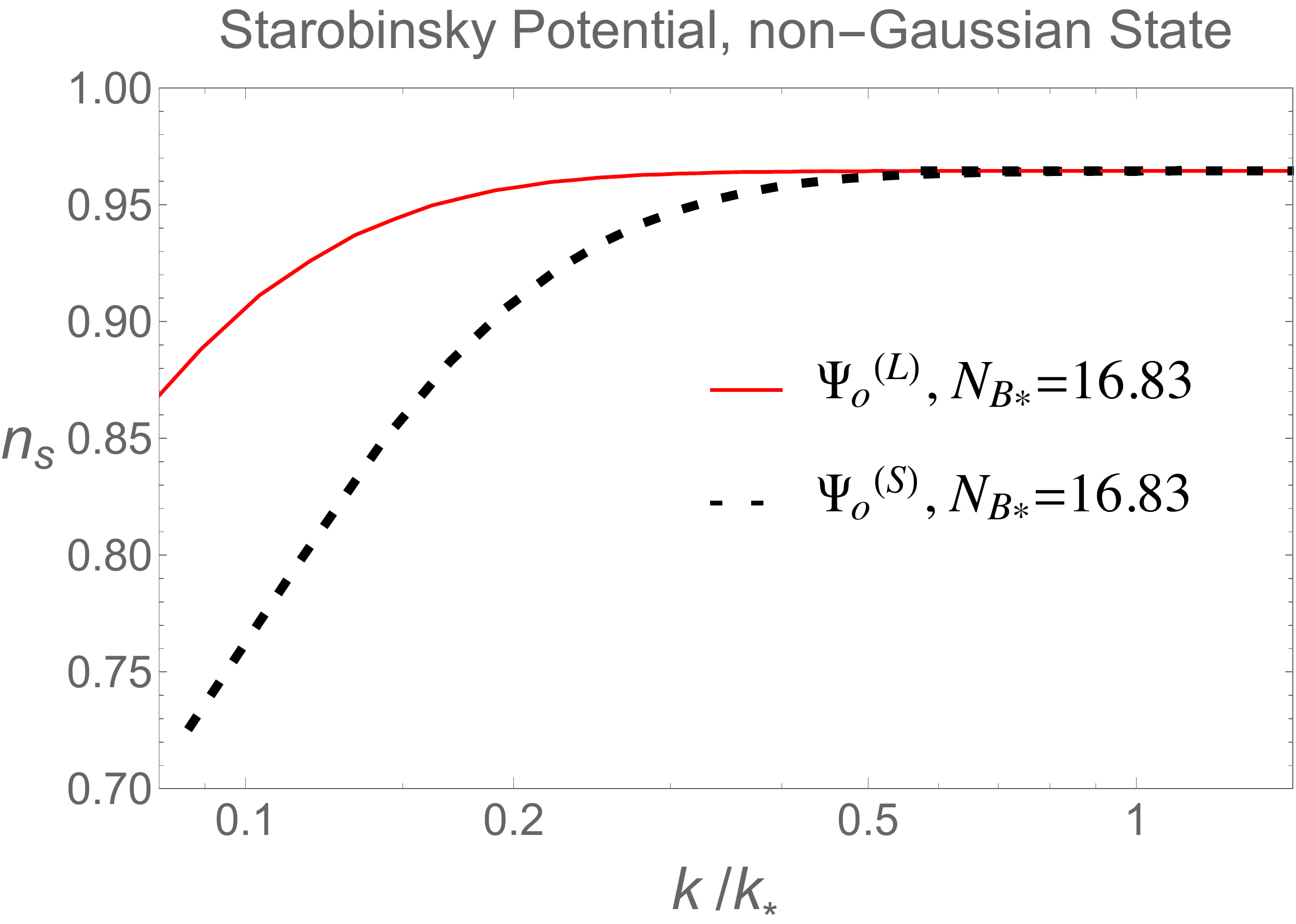}
\caption{{\it Starobinsky potential:} Comparison of observable predictions from a sharply peaked state $\Psi_{o}^{(S)}$ with $\Delta V/V=0.01$ at the bounce and a widely spread state with $\Delta V/V=1$ at the bounce. The widely spread state is} non-Gaussian.\,\,\,  {\it Left panel: Scalar power spectrum. {\it Right panel:} Scalar spectral index. Note that in both plots, the two curves appear to agree if the one resulting from sharply peaked states is shifted slightly to the left. This expectation is borne out by the detailed simulations reported in section \ref{s4.2}.}
\label{fig:powernsNonGaussStarobinsky}
\efig

The physics behind this phenomenon does not depend on whether the state $\Psi_{o}(v,\phi)$ of quantum geometry is widely spread or sharply peaked and our numerical simulations confirmed that this behavior occurs also for widely spread states. Fig. \ref{fig:ppsns1} shows the scalar power spectrum $\mathcal P_{\mathcal{R}}(k)$ for: (i)  the Starobinsky potential;  (ii) a widely spread Gaussian state $\Psi_{o}(v, \phi)$ for the background FLRW geometry,  with $\Delta V/V =1$ at the bounce and $\Nbstar = 16.8$; and, as noted above, (iii) an instantaneous preferred vacuum for scalar perturbations, specified $0.1 \spl$ before the bounce. For the Starobinsky potential, and for the widely spread states under consideration, we have $k_{\star} \approx 16.36\, k_{\, \rm LQC}$, where $k_{\star}$ is the pivot mode used by WMAP (see footnote \ref{1}). The plot shows that new features emerge for $k \lesssim 0.6 k_{\star}  \approx 10\, k_{\,\rm LQC}$. Specifically, for these modes, the power spectrum $P_{\mathcal{R}}(k)$ exhibits rapid oscillations, shown by (blue) crosses. To compare this prediction with observations, one has to introduce appropriately small bins and average the result over them since observations do not yield power at each individual $k$, (and an averaging of this type naturally occurs in going from $P_{\mathcal{R}}(k)$ to the correlation functions $C_{\ell}$ that are generally shown). The red curve shows these averages and it shows that the power spectrum has a greater red tilt for $k \lesssim 0.6 k_{\star} \approx 10 k_{\, \rm LQC}$. Thus, the qualitative behavior is the same as that for sharply peaked states $\Psi_{o}(v,\phi)$ of quantum geometry if one again uses the instantaneous preferred vacuum for scalar perturbations (see, e.g., Fig 4 of \cite{aan3}). The interplay between the ultraviolet and infrared persists also for the widely spread states.

To bring out the relation between observational predictions from sharply peaked states and those with large fluctuations, in Figs. \ref{fig:powernsGaussStarobinsky} and  \ref{fig:powernsNonGaussStarobinsky} we show the two sets of predictions in the same plot.
The two plots in Fig. \ref{fig:powernsGaussStarobinsky} compare the power spectra and spectral indices resulting from a widely spread \emph{Gaussian} state and a sharply peaked state, and the two plots in Fig. \ref{fig:powernsNonGaussStarobinsky} show the same comparison between a widely spread \emph{non-Gaussian} state and the same sharply peaked state. In all plots, the number $\Nbstar$ is kept fixed, $\Nbstar = 16.83$.  What is striking is that there is no qualitative difference between the plots resulting from widely peaked states and sharply peaked states. Because in the sharply peaked state $\Delta V/V = 10^{-2}$ at the bounce, while for the widely spread states we have $\Delta V/V =1$, the two quantum geometries on which the perturbations propagate are very different. One would therefore have expected to see that there would be some new observable features that would clearly stand out. Not only does this not happen but it is apparent that if the plots corresponding  to sharply peaked states were shifted to left a little bit, they would essentially coincide with the plots corresponding to the widely spread states. Simulations performed with intermediate values of $\Delta V/V$ at the bounce show that larger the value of $\Delta V/V$, greater is the required shift.

To understand this feature, let us recall the key difference between the sharply peaked and widely spread states for dynamics of cosmological perturbations. As discussed in section \ref{s2.2}, widely spread states produce dressed metrics that bounce at lower energy densities and lower curvature than sharply peaked ones, and it is the value of the scalar curvature at the bounce that determines $k_{\, \rm LQC}$. Therefore, the LQC scale $k_{\, \rm LQC}^{(S)}$ for sharply peaked states is higher than the scale $k_{\, \rm LQC}^{(L)}$ for states with large relative dispersion. It is this fact that requires us to shift the curves for the sharply peaked states to the left to make the power spectra and spectral indices agree. We will return to this issue in sections \ref{s4.2} and \ref{s5}.\\

\bfig
 \hskip-0.8cm
 \ig[width=0.7\textwidth,height=2.5in]{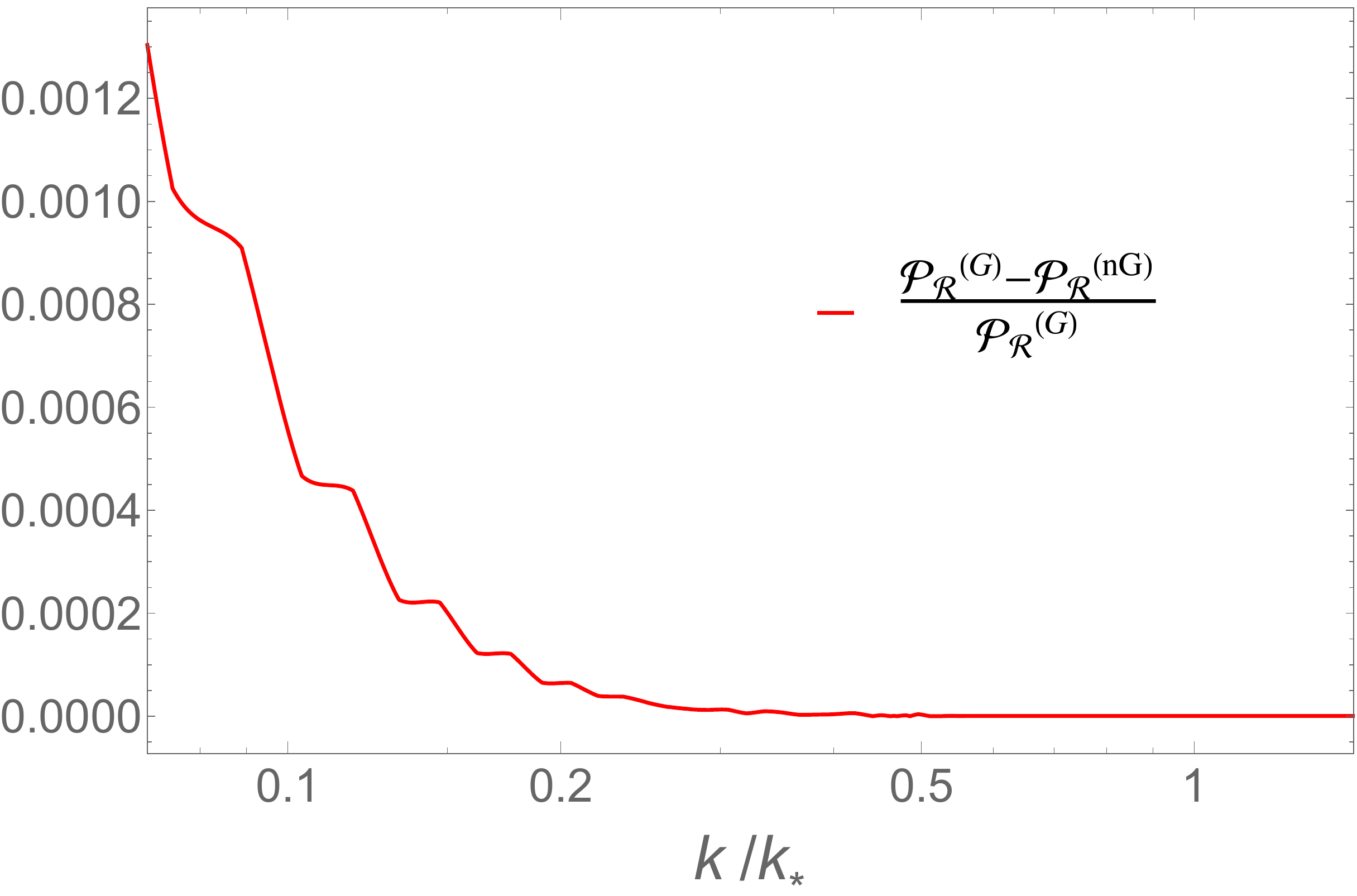}
\caption{Comparison between the binned power spectra for Gaussian and multi-peaked, non-Gaussian states. Both states are widely spread, with $100\%$ relative dispersion at the bounce and $\Nbstar=16.8$, but have very different profiles and therefore quantum fluctuations. In spite of these the relative difference in the power spectrum is only $\lesssim 1.2 \times 10^{-3}$ over the entire observable $k$-range!}
\label{fig:ppsns2}
\efig

Finally, let us ask whether the longest wavelength modes which are excited during the Planck epoch can sense the difference between the Gaussian and non-Gaussian states, which have very different profiles. Fig. \ref{fig:ppsns2} addresses this issue. It compares the power spectra at the end of inflation when the cosmological perturbations propagate on a widely spread \emph{non-Gaussian} quantum geometry state, with that when they propagate on a widely spread \emph{Gaussian}, both with $\Delta V/V =1$ at the bounce. The relative difference has an upper bounded of $0.12\%$ and is in fact less than $0.02\%$ for $k > 0.15 k_{\star}$, i.e. for almost the full range spanned by modes observed by the PLANCK mission. For these long wavelength modes the observational error-bars are rather large --greater than $\sim 5\%$--  whence observations cannot distinguish between the two power spectra.\\ 

Let us summarize. Since the LQC scale $k_{\, \rm LQC}$ at the bounce is  smaller for states with large dispersion than for sharply peaked ones, the conceptual reasoning in the last two paragraphs tells us that the resulting spectra are shifted from each other. On the other hand,  two widely spread states  define the same $k_{\, \rm LQC}$. However, they
have very different profiles throughout the Planck epoch, whence fluctuations in quantum geometry they represent are also quite different. Yet, remarkably, the two power spectra turned out to be observationally indistinguishable.\\ 

\emph{Remark:} We saw in section \ref{s2.3} that the dynamics of cosmological perturbations is sensitive only to the dressed metric $\t{g}_{ab}$ extracted from the given state $\Psi_{o}(v,\phi)$ of the background quantum geometry and not to details of the profile of  $\Psi_{o}(v,\phi)$.  In turn, the observables --$P_{\mathcal{R}}(k)$ and $n_{s}(k)$--  are not sensitive to all the details of the pre-inflationary dynamics of perturbations. As we have just found, primarily they appear to be sensitive only to the LQC scale $k_{\, \rm LQC}$ set by the scalar curvature $R_{\rm B}$ at the bounce. It would have been difficult to anticipate this feature on general grounds.

\subsection{Degeneracy}
\label{s4.2}

Recall that, through its generalized effective metric $\b{g}_{ab}^{\rm (GE)}$, every state $\Psi_{o}$ of the background quantum geometry determines the value $\phib$ of the inflaton $\phi$ at which the bounce occurs, as well as the number $\Nbstar$ of pre-inflationary e-folds. As explained above in section \ref{s4.1}, in LQC the freedom in extending the standard inflationary paradigm to the Planck regime can be encoded in either $\phib$ or $\Nbstar$, but in this paper we use $\Nbstar$ because it has more direct physical meaning. In this sub-section we will show that there is an interesting relation between $\Nbstar$ and the shift, discussed above, that relates the power spectra and the spectral indices of sharply peaked and widely spread states.  

Let begin by recalling the effect a change in $\Nbstar$ has on the power spectrum at the end of inflation. For this, we need to change $\Nbstar$ within the same class of states. Let us therefore consider two sharply peaked states which differ in their values of $\phib$ and hence of $\Nbstar$. For these states, the curvature at the bounce and hence the LQC scale $k_{\, \rm LQC}$ is the same. This means that  the modes whose dynamics is affected by curvature during the Planck epoch have the same range of physical wave-numbers at the bounce for the two sharply peaked states. However, because the number $\Nbstar$ of pre-inflationary e-folds is different, these modes have \emph{different physical wave-numbers} at time $t_{\star}$ that marks the onset of inflation, and hence also in the CMB. Larger the value of $\Nbstar$, greater the value $a_{\rm CMB}$ of the scale factor at the CMB time and hence, for any given physical wave number $k$ at the bounce, smaller the \emph{physical} wave-number at the CMB time. Recall that effects of pre-inflationary LQC dynamics manifest themselves for modes which have $k\lesssim 10\,k_{\, \rm LQC}$ at the bounce, and $k_{\, \rm LQC}$ is the same for the two sharply peaked states under consideration. Therefore, they manifest themselves at different values of $k_{\rm phy}:=k/a_{\rm today}$ in the CMB: Larger the value $\Nbstar$, \emph{lower} the values of $k_{\rm phy}$ at CMB at which quantum gravity effects manifest themselves. 
That is, the power spectrum that emerges from the sharply peaked state $\Psi_{o}(v,\phi)$ with a larger value $\Nbstar$ is shifted to the left relative to the state with a lower value of $\Nbstar$.

\bfig
 \ig[width=0.48\textwidth]{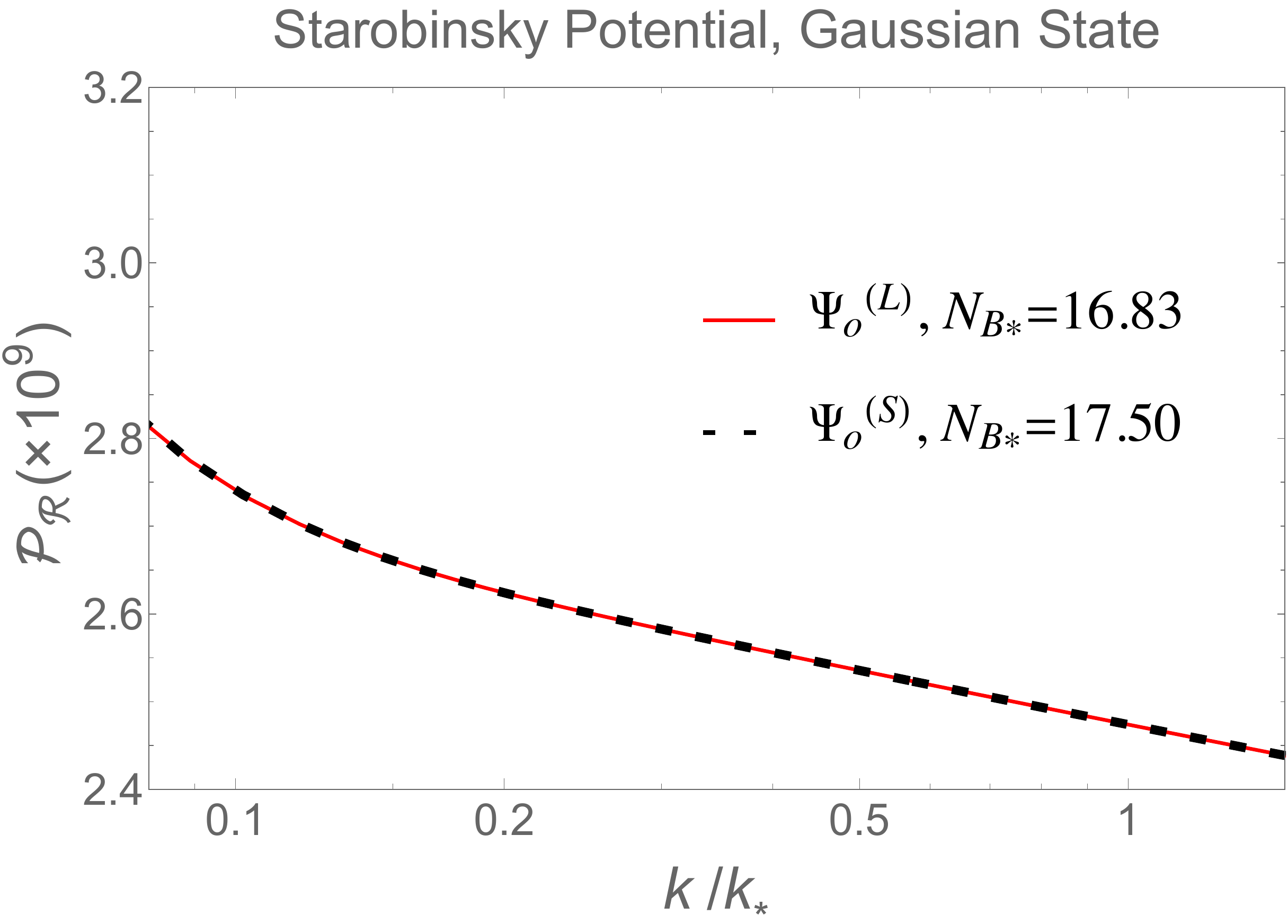}
\hskip0.4cm
 \ig[width=0.48\textwidth]{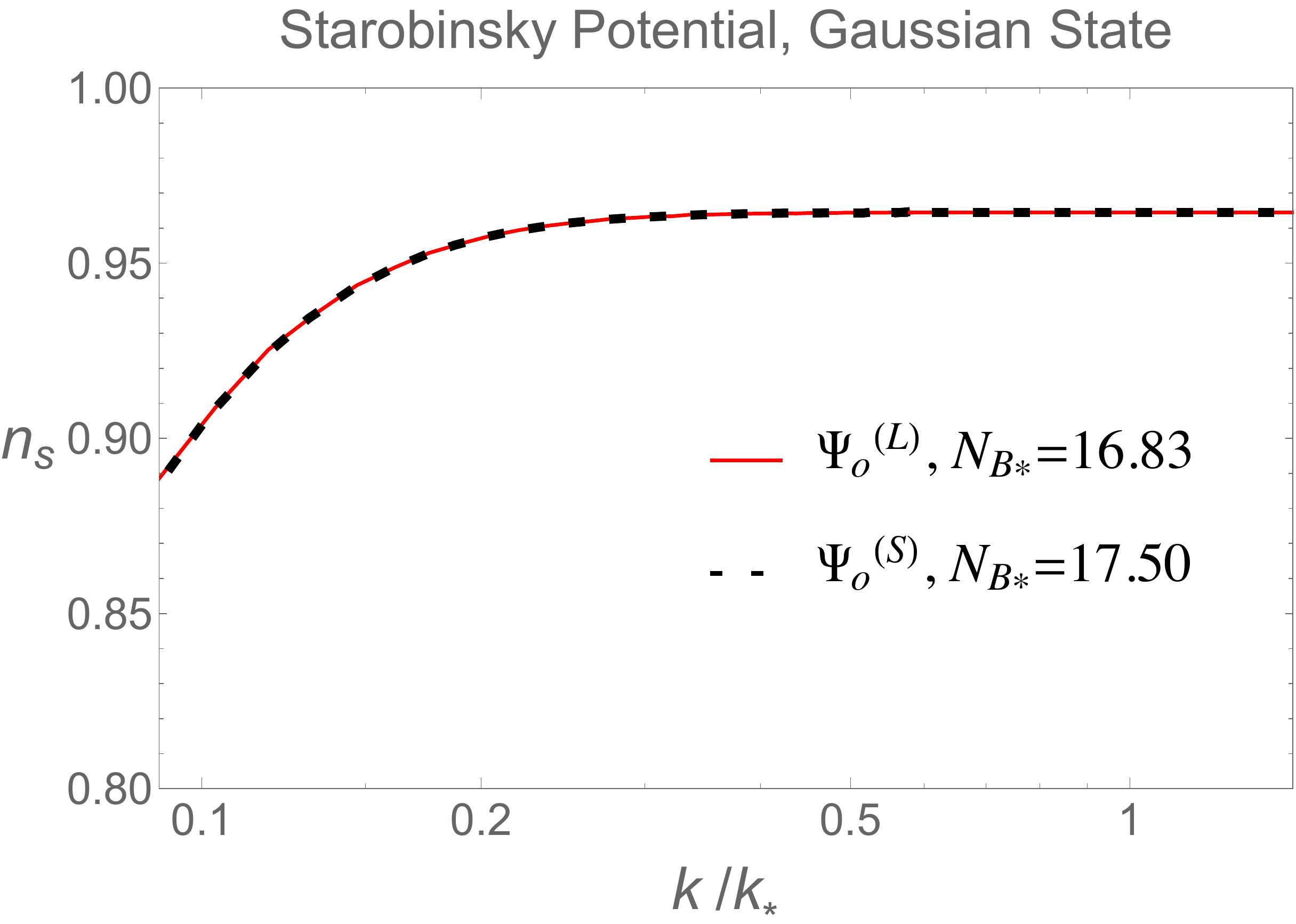}
\caption{Degeneracy between $\Delta V/V$ at the bounce and $\Nbstar$ for the \emph{Starobinsky} potential: widely spread \emph{Gaussian} state. Plots for $\Delta V/V = 1$ and $\Nbstar = 16.83$ for the widely spread state (solid red curve), and $\Delta V/V = 0.01$  and $\Nbstar = 17.50$ for the sharply peaked state (dashed black curve) are virtually indistinguishable. \emph{Left Panel:} Power spectrum $\mathcal{P}_{\mathcal R}(k)$.\,\,  \emph{Right Panel:} Spectral index $n_{s}(k)$.}
\label{fig:degenStarobinskyG}
\efig

\bfig
 \ig[width=0.48\textwidth]{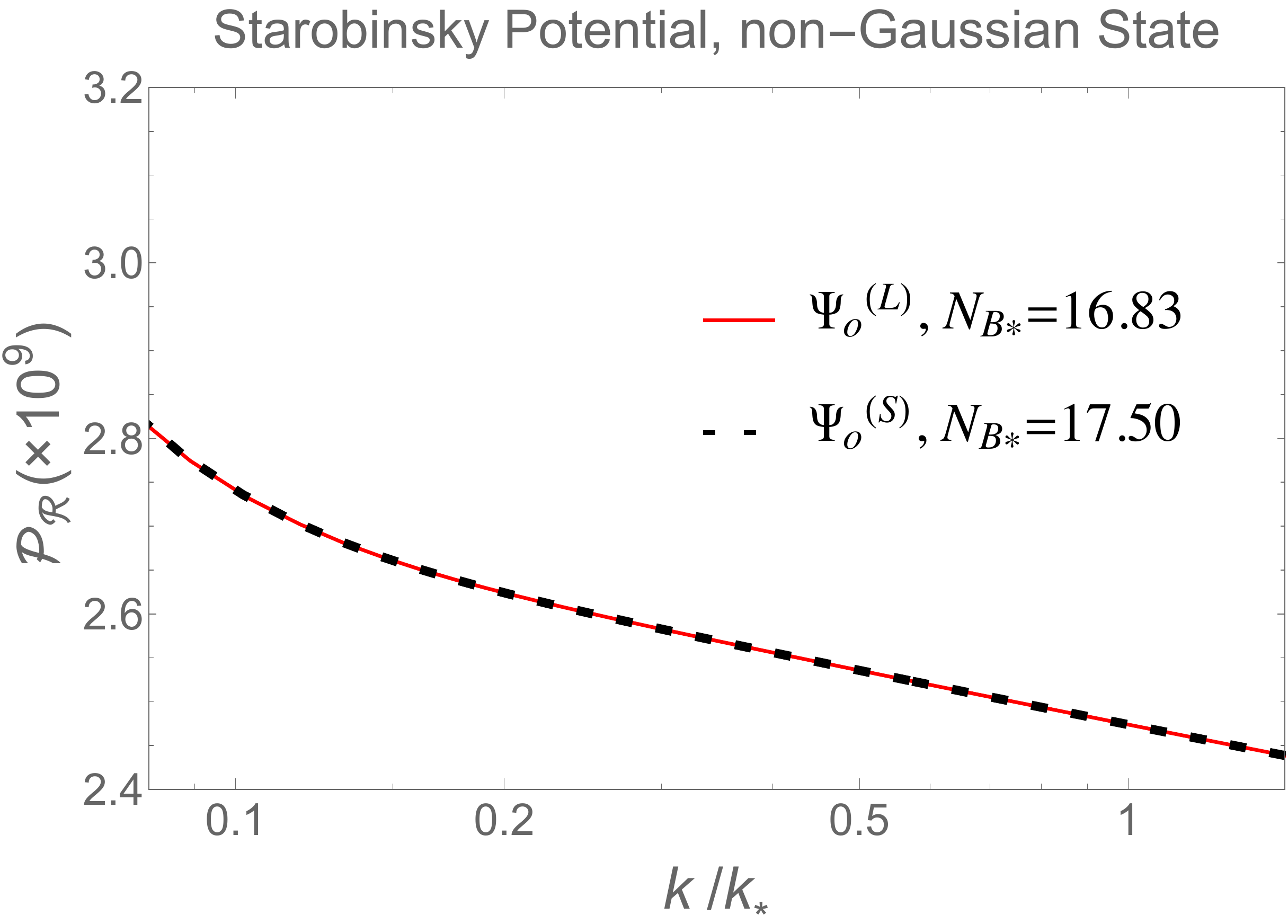}
\hskip0.4cm
 \ig[width=0.48\textwidth]{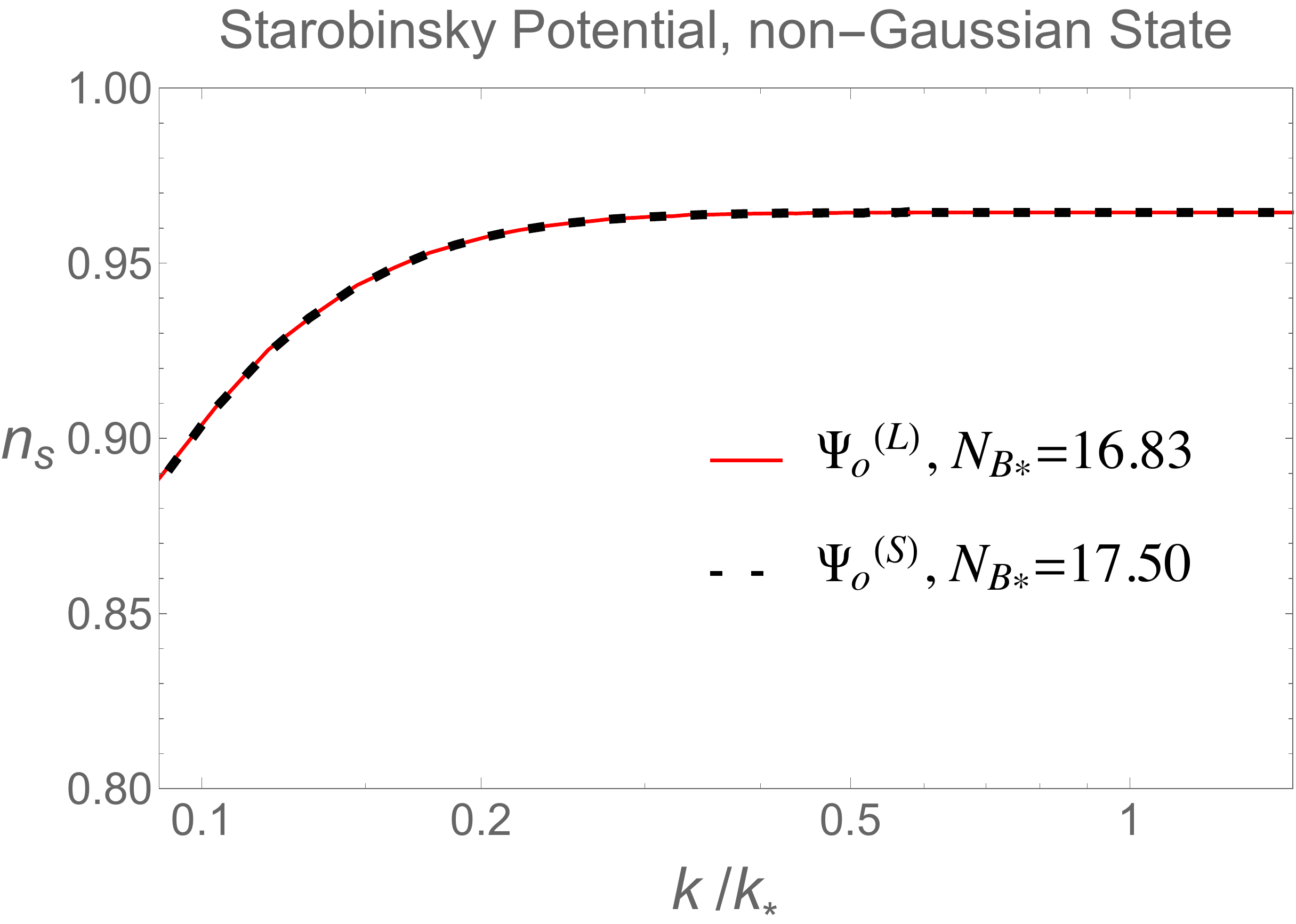}
\caption{Degeneracy between $\Delta V/V$ at the bounce and $\Nbstar$ for the \emph{Starobinsky} potential: widely spread \emph{non-Gaussian} state. Plots for $\Delta V/V = 1$ and $\Nbstar = 16.83$ for the widely spread state (solid red curve), and $\Delta V/V = 0.01$ and $\Nbstar = 17.50$ for the sharply peaked state (dashed black curve) are virtually indistinguishable. \emph{Left Panel:} Power spectrum $\mathcal{P}_{\mathcal R}(k)$.\,\,  \emph{Right Panel:} Spectral index $n_{s}(k)$.}
\label{fig:degenStarobinskynG}
\efig
 
Recall that we found in section \ref{s4.1} that, after a small shift to the left, the power spectrum of a sharply peaked state with a given $\Nbstar$ can be made to coincide with that of a widely spread state with the same $\Nbstar$. Discussion of the last para suggests that the required shift to the left could be brought about simply by considering a sharply peaked state with a slightly larger value of $\Nbstar$. Thus, we are led to the following conjecture:\\  
\vskip0.01cm
\emph{Power spectrum and spectral index arising from a widely spread state $\Psi_{o}^{(L)}$ 
with pre-inflationary e-folds $\Nbstar$ are the same as those arising from a sharply peaked state $\Psi_{o}^{(S)}$ with pre-inflationary e-folds $(\Nbstar +\delta)$ for an appropriately chosen $\delta > 0$.}\\
This statement is surprising because it says that, as far as the CMB observations are concerned, all effects associated with large fluctuations in quantum geometry can be incorporated in the sharply peaked states simply by tweaking the value of the LQC parameter $\Nbstar$. Contrary to one's first intuition, the use of widely spread states with different profiles does not appear to introduce any new ambiguities in the observational predictions of LQC.

Since the statement is surprising, we tested the conjecture using all of possible 8 combinations: (i) widely spread \emph{Gaussian} and  \emph{non-Gaussian} states with $100\%$ relative dispersions at the bounce. As noted in caption of Fig. \ref{fig:wavefun}, in the Planck regime these fluctuation grow to $168\%$ and then plateau;\, (ii) \emph{Starobinsky} and \emph{quadratic} potentials with mass parameters determined using data from the PLANCK mission; and, \, (iii) \emph{power spectrum} $\mathcal{P}_{\mathcal R}(k)$ and \emph{spectral index} $n_{s}(k)$ where $k$ ranges over the entire interval spanned by the PLANCK mission data. Results are plotted in the four figures Figs. \ref{fig:degenStarobinskyG}-\ref{fig:degenQuadnG}

\bfig
 \ig[width=0.48\textwidth]{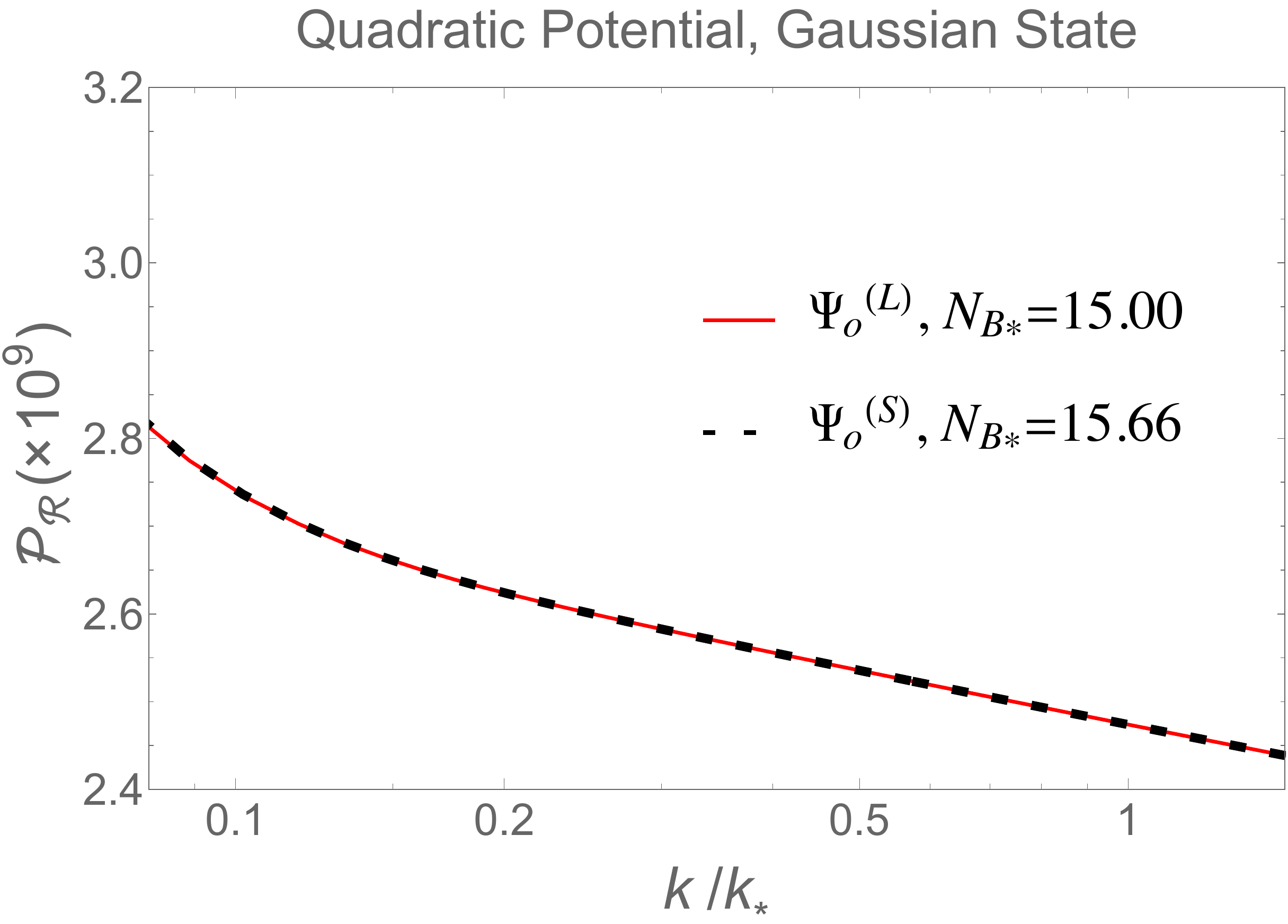}
\hskip0.4cm
 \ig[width=0.48\textwidth]{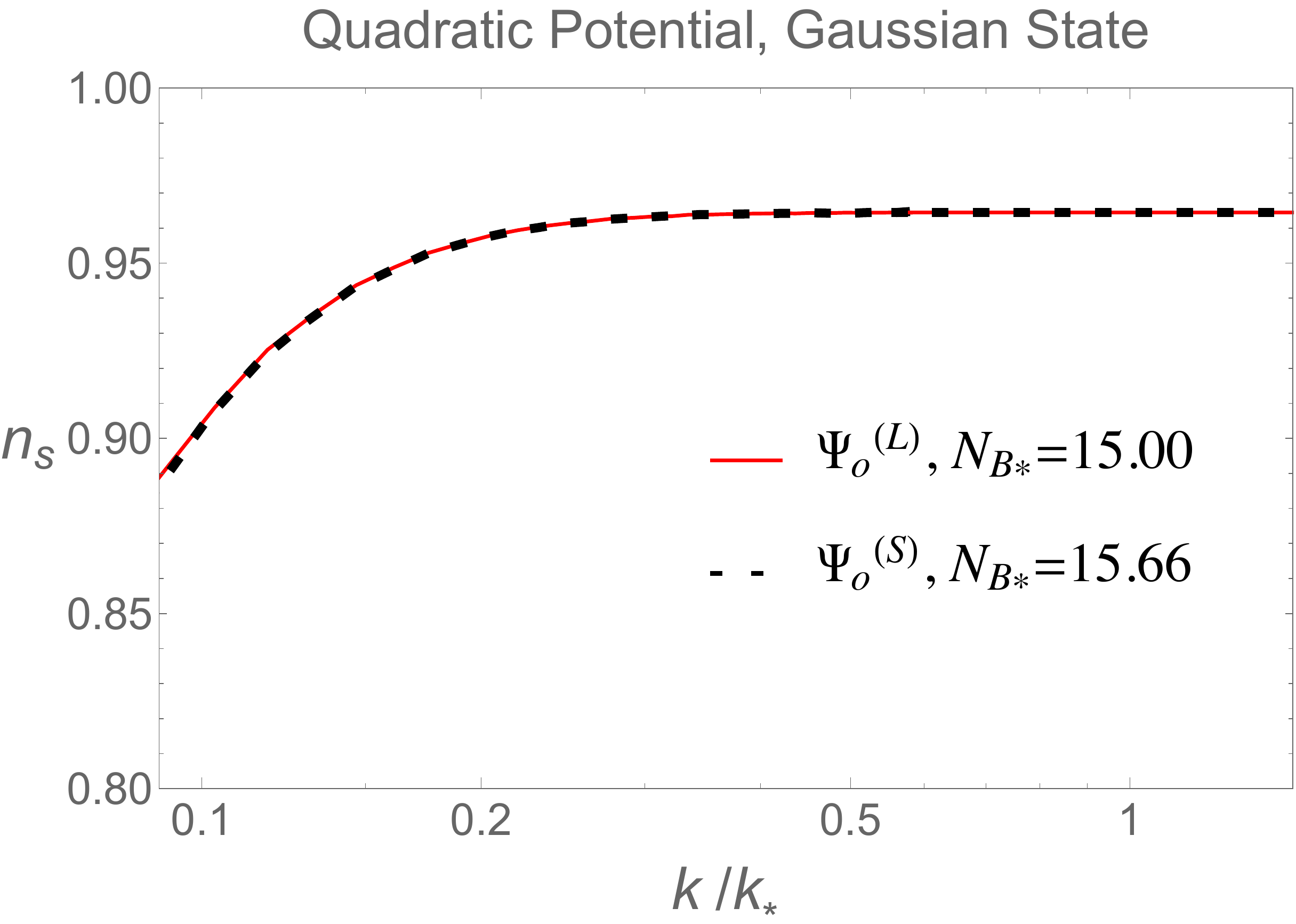}
\caption{Degeneracy between $\Delta V/V$ at the bounce and $\Nbstar$ for the \emph{quadratic} potential: widely spread \emph{Gaussian} state. Plots for $\Delta V/V = 1$ and $\Nbstar = 15.00$ for the widely spread state (solid red curve), and $\Delta V/V = 0.01$ and $\Nbstar = 15.66$ for the sharply peaked state (dashed black curve) are virtually indistinguishable. \emph{Left Panel:} Power spectrum $\mathcal{P}_{\mathcal R}(k)$.\,\,  \emph{Right Panel:} Spectral index $n_{s}(k)$.}
\label{fig:degenQuadG}
\efig

\bfig
 \ig[width=0.48\textwidth]{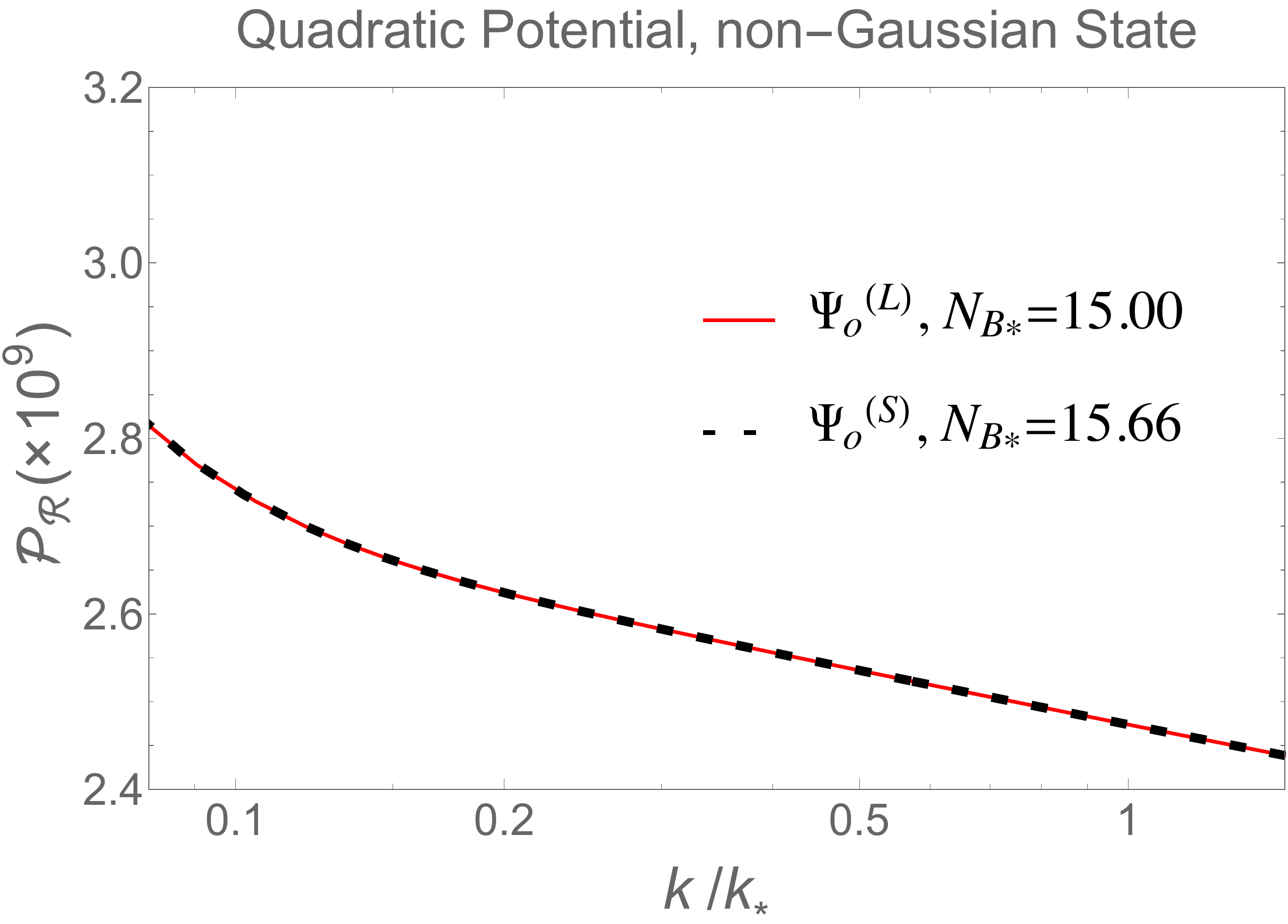}
\hskip0.4cm
 \ig[width=0.48\textwidth]{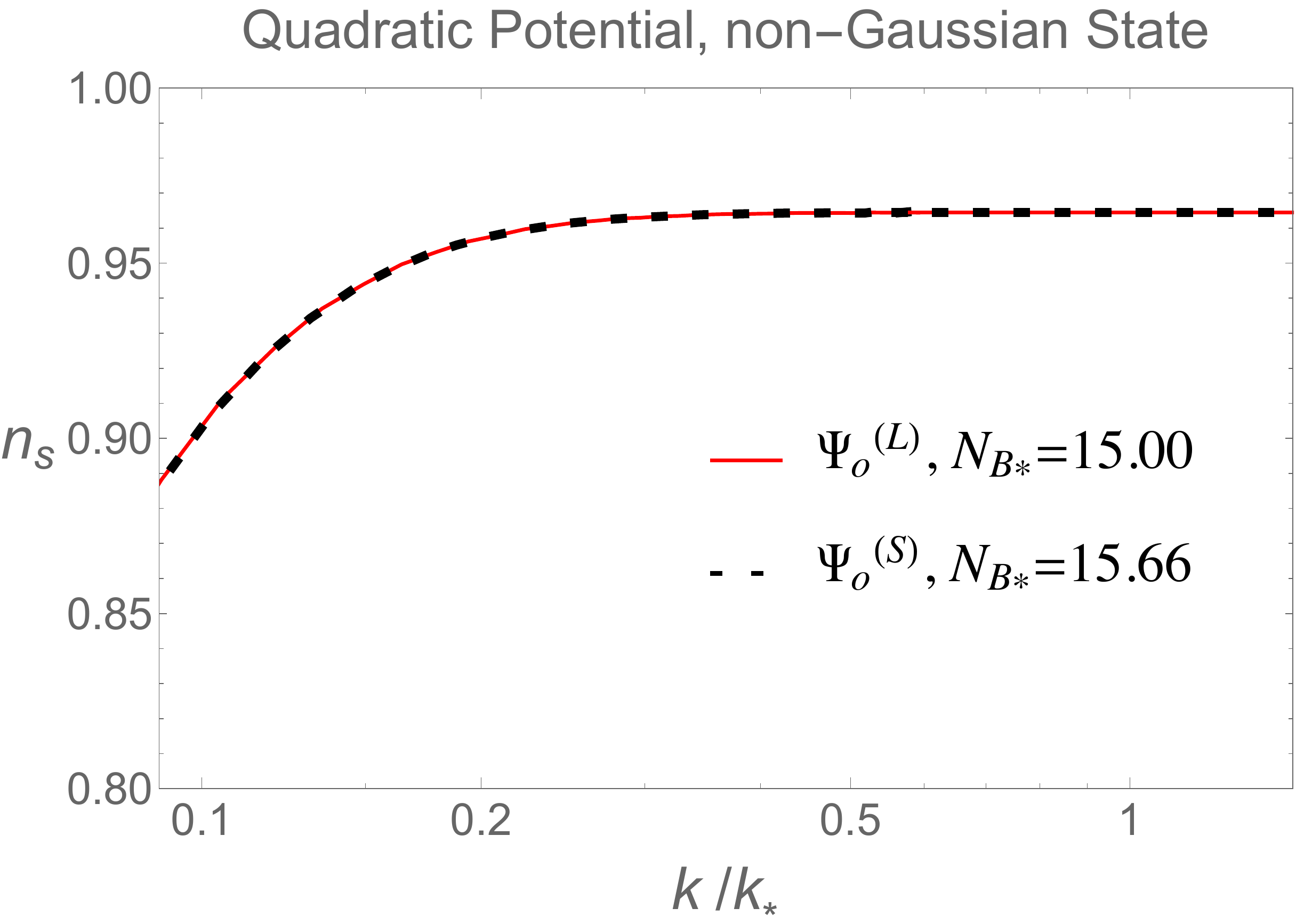}
\caption{Degeneracy between $\Delta V/V$ at the bounce and $\Nbstar$ for the \emph{quadratic} potential: widely spread \emph{non-Gaussian} state. Plots for $\Delta V/V = 1$ and $\Nbstar = 15.00$ for the widely spread state (solid red curve), and $\Delta V/V = 0.01$ and $\Nbstar = 15.66$ for the sharply peaked state (dashed black curve) are virtually indistinguishable. \emph{Left Panel:} Power spectrum $\mathcal{P}_{\mathcal R}(k)$.\,\, \emph{Right Panel:} Spectral index $n_{s}(k)$.}
\label{fig:degenQuadnG}
\efig

In each case the numerical simulation confirmed the conjecture. In Figs.\ \ref{fig:powernsGaussStarobinsky} and \ref{fig:powernsNonGaussStarobinsky}, we used widely spread states --both Gaussian and non-Gaussian-- for which $\Nbstar = 16.83$ for the Starobinsky potential. As discussed in section \ref{s4.1}, with these choices $k_{\,\rm LQC}^{(L)} \approx 0.06\, k_{\star}$ and hence effects of LQC pre-inflationary dynamics manifest themselves for $k \lesssim 10\, k_{\,\rm LQC}^{(L)}\approx 0.6\, k_{\star}$. For the quadratic potential, we were led to consider widely spread states with $\Nbstar=15$ because then the LQC effects manifest themselves also for modes with wave numbers $k \le 0.06\, k_{\star}$. Main results are:

(i) For the Starobinsky potential, we found that both the power spectrum and the spectral index  for the broadly spread states coincide with those of the sharply peaked states with $\Nbstar = 17.50$. Thus we had to \emph{increase} the e-folds of the sharply peaked states, just as our qualitative argument suggested. The surprising feature is that, after the shift, \emph{the agreement is excellent for all observable modes}. In the resolution used in the plots, the two curves are right on the top of each other. Numerical data shows that the maximum relative difference is $\sim 2.3\times 10^{-3}$!

(ii) For the quadratic potential we also found that we needed to increase the number of e-folds for the sharply peaked state. While the widely spread states have $\Nbstar=15$, the sharply peaked states have $\Nbstar=15.66$. Again, the surprising feature is that, after the shift, \emph{the agreement is again excellent for all observable modes} with  the maximum relative difference again of $\sim 2.3\times 10^{-3}$.

The plots show only the range $0.1k_{\star} \lesssim k \lesssim 1.5 k_{\star}$ because: (a) 
corrections due to pre-inflationary LQC dynamics are negligible for modes with larger values of wave numbers $k$; and, (b) modes with $k \lesssim 0.1 k_{\star}$ are not observable because their physical wavelength in the CMB is larger than the radius of the observable universe at the CMB time. \\

The required increase, found numerically by varying the number of e-folds, turned out to be quite small even though the relative dispersion at the bounce in the widely spread states is 100\% and in sharply peaked states is only $1\%$. For the Starobinsky potential, $\Delta\Nbstar = 17.50 -16.83 = 0.67$ while for the quadratic potential $\Delta\Nbstar = 15.66-15.00 = 0.66$. Thus, $\Delta \Nbstar$ is quite close in the two cases. This finding is also in agreement with our general arguments because for both potentials, pre-inflationary LQC dynamics is important for $k \lesssim 10\, k_{\rm LQC}^{(L)} \approx 0.6\, k_{\star}$ for states with large relative dispersions, and for $k \lesssim 10\, k_{\rm LQC}^{(L)} \approx 1.2\, k_{\star}$ for sharply peaked states. We will return to this point in section \ref{s5}.\\

\emph{Remark:} The general, conceptual arguments we introduced in section \ref{s4} clarify why the range of wave numbers for which LQC corrections are significant is the same, first for the two classes of widely spread states we used, and then also for the sharply peaked states with an appropriate shift in the number $\Nbstar$ of pre-inflationary e-folds. However, these arguments do \emph{not} explain why the power spectra and spectral indices agree \emph{in all their details} throughout this range. Before carrying out numerical simulations we did not expect that they would be matched so accurately. 

We studied dynamics with  two different inflaton potentials, allowed highly fluctuating quantum geometries with $100\%$ relative dispersions at the bounce and very different profiles, and compared them with highly peaked states of quantum geometries with only $1\%$ relative dispersions. Still, it is important to bear in mind that our results are restricted to these situations. We do not have any reason to support the view that the results will continue to hold in all circumstances,\, e.g., if we allowed states in which the relative dispersions at the bounce are, say $10^{6}\,\%$, or studied bounces in which the potential energy dominates. Still, we believe that the range of parameters we did allow is large enough to conclude that the LQC results with sharply peaked quantum geometries have a much wider range of applicability than has been anticipated so far in the literature.

\section{Discussion}
\label{s5}

In this paper we investigated the observational consequences of large fluctuations in quantum FLRW geometries in the very early universe.
To carry out this task we made three assumptions:\\
(i) At the bounce, kinetic energy in the inflaton dominates over its potential energy;  \\
(ii) The relative dispersions in quantum geometry are large but not much larger than $100\%$ at the bounce (and $168\%$ in the full quantum gravity regime); and\\
(iii) The back reaction of the \emph{inhomogeneous scalar modes} on the state $\Psi_{o}$ of the background FLRW quantum geometry is negligible. \\
Restriction (i) naturally arises from a general principle (introduced in section III.A of \cite{ag3}) and also supported by purely observational considerations (discussed in section IV.D of \cite{ag3}). Restriction (ii) was made because of computational challenges since, even with more advanced techniques introduced in \cite{dgs2,dgms}, allowing fluctuations much greater than $100\%$  would require significantly more computational resources or time. In any case, from a physical standpoint, the class of fluctuations we allow seem to be more than adequate. On the other hand assumption (iii) could be a true limitation. Therefore, the issue of back reaction is currently being investigated further \cite{agm}.

These assumptions allowed us to make certain approximations. Since the observational error bars on the amplitude and spectral index are a few percent, we made sure that the errors introduced by approximations are $\sim 0.001\%$ or smaller. To check that the results are not not tied to a specific inflaton potential, we carried out all calculations for the two commonly used potentials: the quadratic and the Starobinsky potential. We found that the results are robust.

We began with two types of states $\Psi_{o}$ of the background quantum geometry: Gaussians, and non-Gaussians with multiple peaks at the bounce. At the bounce the two states have the same expectation values and uncertainty in volume, but their profiles --and hence expectation values in higher powers of the scale factor and of other observables-- are very different. In each case, we carried out simulations with different values of relative dispersions $\Delta V/V$ at the bounce, and reported the detailed findings for $100\%$ relative dispersions. (Detailed evolution shows that, as in \cite{ag1}, these dispersions grow further and plateau at about $\sim 168\%$ in the Planck era.)  In spite of large relative dispersions and very different profiles, for observable wave numbers $k$, the final power spectrum $\mathcal{P}_{\mathcal{R}}(k)$ and the spectral index $n_{S}(k)$ turned out to be essentially indistinguishable for the quadratic as well as for the Starobinsky potential. As Fig. \ref{fig:ppsns2} shows, for the Starobinsky potential, the relative difference in $\mathcal{P}_{\mathcal{R}}(k)$ is $\sim 0.1\%$ for wave numbers $k \le 0.2 k_{\star}$ where $k_{\star}$ is the pivot mode, and is much smaller for all higher values of $k$. (Recall that the PLANCK mission data covers wave numbers in the interval $\sim (0.1 k_{\star}, 300k_{\star})$.) Since cosmological perturbations are the only probes we have had into the early universe, current observations do not allow us to distinguish between these very different types of fluctuating background quantum geometries. This finding was unanticipated. 

Furthermore, an even more remarkable feature emerged. Recall first that the freedom in pre-inflationary dynamics of sharply peaked states $\Psi_{o}$ is captured in a new parameter, $\Nbstar$, the number of pre-inflationary e-folds. With a small shift in the value of $\Nbstar$, predictions for $\mathcal{P}_{\mathcal{R}}(k)$ and $n_{S}(k)$ for states $\Psi_{o}^{(L)}$ with large fluctuations can be made to coincide with those obtained from states $\Psi_{o}^{(S)}$ that are sharply peaked during the entire Planck regime. In this sense, states with large fluctuations in quantum geometry in the very early universe cannot be distinguished from the sharply peaked states using CMB observations. Note also that the dynamics of cosmological perturbations is governed entirely by the dressed metric $\t{g}_{ab}$ extracted from $\Psi_{o}$, and we found that $(\t{g}_{ab}, \phi)$ satisfies Einstein's equations very soon after the bounce not only for sharply peaked states but also for those with large relative dispersions (see Fig. \ref{fig:dressed}). Therefore, the fact that successful calculations in the standard inflationary paradigm are based on classical space-time of general relativity does not preclude the possibility that the \emph{physical} quantum geometry $\Psi_{o}$ of our universe had large fluctuations even during inflation. Cosmological perturbations and the CMB inhomogeneities they source are simply insensitive to these fluctuations.

In retrospect, one can account for these findings as follows. The process of calculating $\mathcal{P}_{\mathcal{R}}(k)$ and $n_{S}(k)$ starting from any given quantum geometry state $\Psi_{o}$ involves two steps, each of which `filters out' certain aspects of the detailed information contained in $\Psi_{o}$. First, one has to calculate the dressed effective metric $\t{g}_{ab}$ which is all that the dynamics of cosmological perturbations depends on. Distinct quantum states $\Psi_{o}$ can give rise to the same $\t{g}_{ab}$ because the expressions (\ref{dresseda}) and (\ref{dressedeta}) of $\t{a}$ and $\t\eta$ depend only on just a few specific characteristics of the state $\Psi_{o}$. In the second step one extracts the observationally relevant quantities $\mathcal{P}_{\mathcal{R}}(k)$ and $n_{S}(k)$ from the dynamics of the scalar modes $q_{k}(\t\eta)$ on the background geometry  of $\t{g}_{ab}$. Again, distinct  $q_{k}(\t\eta)$ can give rise to the same power spectrum and spectral index at the end of inflation. Therefore it is possible for two very different quantum geometries $\Psi_{o}$ to lead to the same $\mathcal{P}_{\mathcal{R}}(k)$ and $n_{S}(k)$. 

What is the situation with the two classes of states with wide dispersions and \emph{very}  different profiles at the bounce?  Let us first note that we need to calculate $\t{a}$ and $\t\eta$ from the full wave function $\Psi_{o}$ \emph{only} in the first $\sim 10\, \spl$, because after this period the matter density and curvature are $\lesssim 10^{-4}$ in Planck units and  $(\t{g}_{ab}, \phi)$ satisfy Einstein's equations to an excellent approximation \cite{ag3}. Fig. \ref{fig:dressed-comparison} shows that, during this Planck epoch, the Gaussian and the non-Gaussian states with large fluctuations lead to \emph{very} similar dressed effective metrics: the maximum relative difference between $\t{a}, \t\eta$ in the first $\sim 12\, \spl$ is $\lesssim 0.2\%$ and $\lesssim 0.4\% $ respectively. Thus, already at the level of the first `filter' much of the distinction between the Gaussian and the non-Gaussian states with very different profiles is removed. We did not anticipate this drastic simplification because the expressions (\ref{dresseda}) and (\ref{dressedeta}) of $\t{a}$ and $\t\eta$ are rather complicated. A priori one would expect them to be sensitive to detailed features of the profiles of the wave functions $\Psi_{o}$. But calculations revealed that this is not the case: the specific combination of operators that feature in Eqs. (\ref{dresseda}) and (\ref{dressedeta}) is such that the resulting dressed effective quantities $\t{a}$ and $\t\eta$ are largely insensitive to the differences. Finally, as Fig. \ref{fig:ppsns2} shows, the relative differences between $\mathcal{P}_{\mathcal{R}}(k)$ (or $n_{S}(k)$) is $\lesssim 0.1\%$ for the two types of widely spread states we considered.

Similar considerations shed light as to why the power spectrum and spectral index obtained from \emph{widely spread} states are essentially indistinguishable from those resulting from \emph{sharply peaked} states with a slightly larger number of pre-inflationary e-folds. The radius of curvature $\rb$ of the dressed metric $\t{g}_{ab}$ at the bounce defines a LQC length scale. Fourier modes of cosmological perturbations which have physical wavelength $\lambdap \ll \rb$ do not experience curvature during the pre-inflationary epoch. Therefore LQC observational predictions for these modes are indistinguishable from those of standard inflation. Situation is very different for modes with $\lambdap \gtrsim 0.1\,\rb$. These modes do experience curvature during the pre-inflationary LQC evolution, whence their power spectrum at the end of inflation is different from standard inflation. (For details, see section 4.1 of \cite{aan3}.) Thus, the key question is: How does the radius of curvature $\rb$ for states that are widely spread at the bounce compare to that of sharply peaked states. Now, for states $\Psi_{o}^{(L)}$ with large dispersions, the bounce occurs at a smaller matter density and curvature than for sharply peaked states $\Psi_{o}^{(S)}$, whence $\rbL > \rbS$. Therefore, if the number of pre-inflationary e-folds $\Nbstar$ for the two states were the same, then their predicted power spectrum would be different: LQC effects would start appearing \emph{in CMB} at shorter wavelengths for the sharply peaked states than for the widely spread states. However, if we consider sharply peaked states for which the e-folds $\Nbstar^{(S)}$ is appropriately larger, then the LQC effects will manifest themselves at the same wavelength \emph{in CMB} as for the widely spread states. Since the physical wavelengths increase linearly with the scale factor, it is clear that the appropriate number is $\Nbstar^{(S)} - \Nbstar^{(L)} = \ln \,\big(\rbL/\rbS\big)$. The widely spread states considered in this paper have relative dispersions of $100\%$ at the bounce while the sharply peaked states have relative dispersions of $1\%$. As a consequence \cite{ag1}, we have $\big(\rbL/ \rbS \big) \approx  1.95$. Therefore if we set $\Nbstar^{(S)} - \Nbstar^{(L)}= \ln 1.95 \approx 0.67$, for these two classes of states, the LQC effects will become manifest at the same range of wavelengths in the CMB. This is precisely the necessary shift we found (see Figs. \ref{fig:degenStarobinskyG}-\ref{fig:degenQuadnG}). Numerical results  showed that the shift is insensitive to whether we use  Gaussian or non-Gaussian widely spread states, and to whether we use the quadratic or the Starobinsky potential, just as our general argument implies. But the general argument has its limitations. In terms of comoving wavenumbers used in the plots, it explains that, after the shift in the number of e-folds, $\mathcal{P}_{\mathcal{R}}(k)$ and $n_{S}(k)$ calculated from sharply peaked and widely spread states will agree for all wave numbers $k$ with $(k/a_{\rm B}) \geq 10\,(\rb)^{-1}$ (because they both agree with predictions of standard inflation in this range). But it does not explain why the predictions from sharply peaked and widely spread states continue to agree \emph{in all their details} so well for smaller wavenumbers.. Explicit numerical simulations were necessary to discover that this is the case.

Let us summarize. In the 1980s, it was often suggested that space-time was irregular at all scales during the Planck era, representing a thermal foam, or a fractal structure (see, e.g., \cite{davies}). On the other hand, investigations in LQC over the past decade showed that while the physical Hilbert space $\Hp$ of quantum geometry does allow states with large fluctuations in the Planck era, it also admits states that remain sharply peaked around an effective trajectory throughout the Planck epoch. Furthermore, it suffices to restrict oneself to such states to account not only for various features observed in the CMB, but also to search for observable signatures of Planck scale physics (see, e.g., \cite{aan3,am,agulloassym,ag3,agullocorichi,abrev}). Results obtained in this paper strengthen this view. For, even if the `real' physical state of quantum geometry $\Psi_{o}$ does have large fluctuations in the early universe, under the three assumptions spelled out at the beginning of this section, it is possible to absorb their effect on the dynamics of cosmological perturbations by continuing to work with sharply peaked states of quantum geometry and simply shifting the number $\Nbstar$ of pre-inflationary e-folds. Since the only probes into the very early universe we currently know of are cosmological perturbations, large fluctuations in geometry could well continue all the way to the end of inflation. If the assumption (iii) above can be justified, or the analysis can be generalized to incorporate the back reaction of cosmological perturbations \cite{agm}, we would conclude that whatever the exotic behavior of quantum geometry may be in the early universe, it will not be reflected in CMB, and hence in the large scale structure of the universe. In this sense, it would remain invisible to us. The pre-inflationary dynamics of quantum states  $\Psi_{o}$ of FLRW geometry \emph{will lead} to observable deviations from standard inflation in CMB at the largest angular scales \cite{aan3,am,agulloassym,ag3}, but these signatures will not be sensitive to the details of the fluctuations in $\Psi_{o}$.

\section*{Acknowledgments}
We would like to thank Javier Olmedo, Jorge Pullin and especially Param Singh for discussions. This work was supported by the NSF grants PHY-1403943, PHY-1505411, PHY-1552603 and the Eberly research funds of Penn State. It used the Extreme Science and Engineering Discovery Environment (XSEDE), which is supported by National Science Foundation grant number ACI-1053575.


\end{document}